\documentclass[aps,pre,twocolumn,groupedaddress,showpacs,floatfix,superscriptaddress]{revtex4-1}
\usepackage[utf8]{inputenc}
\usepackage[T1]{fontenc}
\usepackage{graphicx}
\usepackage{amssymb}
\usepackage{nicefrac}
\usepackage{xcolor}
\usepackage{here}
\usepackage{afterpage}
\usepackage{times}

\usepackage[colorinlistoftodos,prependcaption,textsize=small]{todonotes}
\usepackage{amsmath}
\DeclareMathOperator{\sign}{sign}

\usepackage{hyperref}

\hypersetup{
  colorlinks=true,
citecolor=black,
linkcolor=black,
urlcolor=black
  }

\newcommand{\egn}[1]{\textcolor{black}{#1}}
\newcommand{\bdt}[1]{\textcolor{black}{#1}}
\newcommand{\kc}[1]{\textcolor{black}{#1}}

\begin{document}

\title{Nonlinear friction in underdamped anharmonic stochastic oscillators}

\author{Karol Capa{\l}a}
\email{karol@th.if.uj.edu.pl} \affiliation{
Institute of Theoretical Physics and Mark Kac Center for Complex Systems
Research, Jagiellonian University, ul. St. {\L}ojasiewicza 11,
30--348 Krak\'ow, Poland}

\author{Bart{\l}omiej Dybiec}
\email{bartek@th.if.uj.edu.pl} \affiliation{
Institute of Theoretical Physics and Mark Kac Center for Complex Systems
Research, Jagiellonian University, ul. St. {\L}ojasiewicza 11,
30--348 Krak\'ow, Poland}

\author{Ewa Gudowska-Nowak}
\email{ewa.gudowska-nowak@uj.edu.pl} \affiliation{
Institute of Theoretical Physics and Mark Kac Center for Complex Systems
Research, Jagiellonian University, ul. St. {\L}ojasiewicza 11,
30--348 Krak\'ow, Poland}

\date{\today}

\begin{abstract}

\bdt{Non-equilibrium} stationary states of overdamped anharmonic stochastic oscillators driven by L\'evy noise are typically multimodal.
The very same situation is recorded for an underdamped L\'evy noise driven motion in single-well potentials with linear friction.
Within current manuscript we relax the assumption that the friction experienced by a particle is linear.
Using computer simulations, we study underdamped motions in single-well potentials in the regime of nonlinear friction.
We demonstrate that it is relatively easy to observe multimodality in the velocity distribution as it is determined by the friction itself and it is the same as the multimodality in the overdamped case with the analogous deterministic force.
Contrary to the velocity marginal density, it is more difficult to induce multimodality in the position.
Nevertheless, for a fine-tuned nonlinear friction, the spatial multimodality can be recorded.
\end{abstract}


\maketitle


\textbf{Properties of dynamical systems driven by L\'evy noise are very different from their Gaussian white noise driven counterparts.
For instance, in the overdamped regime, in order to bound L\'evy flights the potential well needs to be steep enough.
Surprisingly, for single-well potentials steeper than parabolic \bdt{non-equilibrium} stationary states (NESS) are bimodal.
Properties of overdamped systems are better explored than properties of underdamped systems.
Therefore, within the current manuscript,  we continue studies on the L\'evy noise driven dynamics in underdamped systems in the regime of linear and nonlinear friction.
Using \bdt{heuristic} arguments and numerical simulations, we explore the role of underlying assumptions and study NESS properties along with conditions for their existence.
We demonstrate that, in the underdamped regime, it is easy to \egn{induce  bimodality in the velocity probability distribution function (PDF), because the phenomenon is determined solely by the form of velocity-dependent friction}.
Contrary to the multimodality in the velocity, nonlinear friction typically results in unimodal marginal position distributions.
Nevertheless, for \egn{suitably predefined nonlinear friction, spatially-separated modes in PDF can be generated.}
Furthermore, superlinear friction \egn{is shown to weaken} the condition on the steepness of single-well potentials which are capable of bounding underdamped L\'evy noise driven motions.
}

\section{Introduction and model}

Motion of a stochastic particle in the presence of a conservative force, damping and thermal fluctuations is conveniently described by the Langevin equation
\begin{equation}
\ddot{x}(t)= - \gamma \dot{x}(t) -V'(x) + \zeta (t),
\label{eq:full-langevin}
\end{equation}
where $\gamma$ ($\gamma>0$) is the damping, $\zeta(t)$ stands for the noise and the unit mass term $m=1$ is assumed. By attributing thermal origin to the fluctuations of the stochastic force, $\zeta(t)$ can be modeled as Gaussian and white with $\langle \zeta(t) \rangle=0$ and $\langle \zeta(t)\zeta(s) \rangle = \gamma \sigma ^2 \delta(t-s)$, as damping $\gamma$ and strength of fluctuations $\sigma$ are then connected by a celebrated Einstein's relation \cite{risken1996fokker,sekimoto2010stochastic}.
Presence of noise randomizes trajectories $(x(t),\dot{x}(t)=v(t))$ making them different even for the same initial conditions.
Consequently, an ensemble of particles immersed in a given $(x(0),v(0))$  point starts to diffuse. Erratic trajectories of the ensemble do not allow for the measurement of the particle's velocity, while it is possible to measure the mean-square displacement from the initial position and show that it is growing linearly in time.

The time evolution of the full probability density associated with Eq.~(\ref{eq:full-langevin}) is described by the Kramers equation \cite{risken1996fokker}
\begin{equation}
 \frac{\partial P }{\partial t}=\left[- v \frac{\partial}{\partial x} +\frac{\partial}{\partial v}\left( \gamma v + V'(x) \right) +\gamma \sigma^{2} \frac{\partial^2}{\partial v^2} \right]P,
 \label{eq:kk}
\end{equation}
\normalsize
where $P=P(x,v,t|x_0,v_0,t_0)$.

Stationary states for the model described by Eq.~(\ref{eq:full-langevin}) and associated long-time solutions to the diffusion equation~(\ref{eq:kk}) exist for any confining potential $V(x)$ increasing to infinity as $|x|\to\infty$. 
More importantly, they are given by the equilibrium  Boltzmann-Gibbs distribution, thus establishing a relation with thermodynamics:
\begin{equation}
    P(x,v) \propto \exp\left[ - \frac{1}{\sigma^2} \left(\frac{v^2}{2} + V(x) \right) \right].
    \label{eq:bg}
\end{equation}
\bdt{The form of the stationary density given by Eq.~(\ref{eq:bg}) clearly indicates that velocity and position are statistically independent. 
Moreover, in the system described by Eq.~(\ref{eq:full-langevin}) with the Gaussian white noise, the condition of detailed balance is fulfilled \cite{reichl1998,huang1963}.
These two, important equilibrium properties are not satisfied under action of L\'evy noises\cite{garbaczewski2011levy,sokolov2011,kusmierz2013onsagers,kusmierz2016breaking,kusmierz2018thermodynamics}.
}

Within the current manuscript, using methods of stochastic dynamics, we will be exploring properties of non-equilibrium stationary states \bdt{(NESS)} for models described by the full, underdamped Langevin equation in the regime of nonlinear dissipative force. Models of that type refer to non-equilibrium cases where the friction is not a constant but a function of the velocities $\gamma=T(v)$  and the Einstein relation is no longer fulfilled. Interesting applications of models of Brownian motion with nonlinear friction have been addressed in various fields: mechanical devices like microspeakers and vibration isolation systems and energy harvesters \cite{Elliott}, self-organized systems exhibiting sustained oscillations \cite{klimontovich1994}, description of motion of charged particles in plasma \cite{Dunkel} or active Brownian motion models of biological motors \cite{egn2009}. 

To start with, let us briefly recollect a special limit of Eq.~(\ref{eq:full-langevin}) with the strong damping.
At strong friction the velocity can be adiabatically eliminated \cite{gardiner2009} from Eq.~(\ref{eq:full-langevin}) resulting in the overdamped Langevin equation
\begin{equation}
     \gamma \dot{x}(t) = -V'(x) + \zeta (t).
    \label{eq:langevin}
\end{equation}
The motion described by Eq.~(\ref{eq:langevin}) is spatially diffusive and fully characterized by the position only. 
The time evolution of the probability density $P(x,t|x_0,t_0)=\langle \delta(x-x(t))\rangle\equiv P$ fulfills the Smoluchowski-Fokker-Planck equation
 \begin{equation}
 \frac{\partial P}{\partial t} = \frac{1}{\gamma}
 \frac{\partial}{\partial x} \left[  -V'(x) + \sigma^2 \frac{\partial}{\partial x}  \right] P,
 \label{eq:sfp}
 \end{equation}
with the stationary solution given again by the Boltzmann-Gibbs form
\begin{equation}
P(x) \propto \exp\left[-\frac{V(x)}{\sigma^2}\right].    
\end{equation}

In more general realms the noise $\zeta(t)$ does not need to be Gaussian. For example it can be of the L\'evy, $\alpha$-stable type.
The symmetric L\'evy noise is the formal time derivative of the symmetric $\alpha$-stable motion $L(t)$, see Ref.~\onlinecite{janicki1994b}, whose characteristic function $\phi(k)=\left\langle \exp[ikL(t)] \right\rangle$ is
\begin{equation}
 \phi(k)=\exp\left[ -  t \sigma^\alpha |k|^\alpha \right]
 \label{eq:fcharakt}
\end{equation}
with the parameter $\sigma$ scaling the strength of fluctuations.
Following this definition $\zeta(t)$ is a symmetric, Markov  $\alpha$-stable noise which turns into a standard Gaussian form for $\alpha=2$. However, unlike  standard Brownian motions for which the mean-square displacement (MSD) grows linearly in time, the dispersion of the position in the L\'evy motion (cf. Eq.~(\ref{eq:langevin})  with $V(x)\equiv 0$) diverges and the width of the resulting asymptotic L\'evy (super)-diffusion  must be characterized  by some  fractional moments \cite{janicki1994,samorodnitsky1994,dubkov2008} or the interquantile distance.
Symmetric $\alpha$-stable densities are unimodal probability densities which for $\alpha<2$ exhibit a power-law asymptotics with tails decaying as $ |x|^{-(\alpha+1)}$.
\bdt{Moreover, for L\'evy noise driven systems, the condition of detailed balance is not satisfied \cite{garbaczewski2011levy,kusmierz2016breaking}}

In case of motions described by the Langevin equations and perturbed by a generalized L\'evy noise, the associated diffusion equations (\ref{eq:kk}) and (\ref{eq:sfp}) become fractional Smoluchowski-Fokker-Planck or Kramers equations \cite{podlubny1999,chechkin2000linear,dubkov2008}.
In Eq.~(\ref{eq:kk}), $\partial^2/\partial v^2$ is then replaced by $\partial^\alpha/\partial |v|^\alpha$,  see Ref.~\onlinecite{lutz2001fractional}, while in  Eq.~(\ref{eq:sfp})
$\partial^2/\partial x^2$ is exchanged with $\partial^\alpha/\partial |x|^\alpha$, see Ref.~\onlinecite{yanovsky2000,schertzer2001}.
The Riesz-Weil $\partial^\alpha/\partial |x|^\alpha$ fractional derivative \cite{podlubny1999,samko1993} is  defined via the Fourier transform $
 \mathcal{F}_k\left( \frac{\partial^\alpha f(x)}{\partial |x|^\alpha} \right)=-|k|^\alpha \mathcal{F}_k\left(f(x)\right).$
 
 The significant differences in statistical properties  of systems driven by non-Gaussian L\'evy  fluctuations, and in particular divergence of the second moment, imply lack of  a simple Einstein's fluctuation-dissipation relation between  fluctuations' strength and  magnitude of  dissipation \cite{jespersen1999,kusmierz2014,kusmierz2016breaking,kusmierz2018thermodynamics}. Accordingly, in Eqs.~(\ref{eq:full-langevin}) and~(\ref{eq:kk}), the damping $\gamma$ and  the noise strength $\sigma$ have to be interpreted as independent parameters. Consequently, for $\alpha<2$, in Eq.~(\ref{eq:kk})    $\gamma \sigma^2 \partial^2/\partial v^2 \to \sigma^\alpha \partial^\alpha/\partial |v|^\alpha$, while in Eq.~(\ref{eq:sfp})
 $\sigma^2 \partial^2/\partial x^2 \to \sigma^\alpha \partial^\alpha/\partial |x|^\alpha$.

L\'evy processes have been  massively studied on theoretical and numerical levels \cite{metzler2000,barkai2001,anh2003,brockmann2002,chechkin2006,jespersen1999,yanovsky2000,schertzer2001}.
Because of significant likelihood of observation of long jumps, L\'evy noises and L\'evy statistics can be successfully applied to description of catastrophic events like economic crises \cite{stanley1986,mantegna2000}, outburst of epidemics \cite{newman1999} or climate changes \cite{ditlevsen1999}. The significant number of observations confirms presence of non-Gaussian fluctuations in the variety of complex dynamical systems
and experimental setups. Among others, L\'evy flights have been recorded in financial time series \cite{bouchaud1990}, rotating flows \cite{solomon1994}, superdiffusion of micellar systems \cite{bouchaud1991}, transmission of light in polidispersive materials \cite{barthelemy2008},  photon scattering in hot atomic vapors \cite{mercadier2009levyflights}, dispersal patterns of humans and animals \cite{brockmann2006,sims2008}, laser cooling \cite{cohen1990,barkai2014}, 
 gaze dynamics \cite{amor2016} and search strategies \cite{shlesinger1986,reynolds2009}.

In the overdamped regime described by Eqs.~(\ref{eq:langevin}) and~(\ref{eq:sfp}) and under the action of an harmonic potential $V(x)=x^2/2$, \bdt{NESS} in the presence of additive  L\'evy noises are given by the rescaled $\alpha$-stable density with the same stability index $\alpha$ as the noise \cite{jespersen1999,chechkin2002,chechkin2003,dybiec2007d}. This is a natural consequence of action of the deterministic linear force and the generalized central limit theorem \cite{feller1968}.
In a more general potential wells the turnover from unimodal to bimodal \bdt{non-equilibrium} stationary probability densities occurs \cite{chechkin2004}. As an exemplary case, we refer to L\'evy flights in the potential $V(x)=x^4/4$, when the Langevin equation takes the following form
 \begin{equation}
\gamma \dot{x}(t) = -x^3(t) + \zeta (t).
    \label{eq:langevin_n4}
\end{equation}
For the L\'evy noise with $\alpha=1$ (Cauchy noise), the \bdt{NESS} of the system can be readily derived \cite{chechkin2002,chechkin2003,chechkin2004,chechkin2006,chechkin2008introduction} and is given by
\begin{equation}
    P_{\alpha=1}(x)=\frac{1}{\pi\sigma^{\nicefrac{1}{3}}\left[(x/\sigma^{\nicefrac{1}{3}})^4-(x/\sigma^{\nicefrac{1}{3}})^2+1\right]}
    \label{eq:stationary-n4}.
\end{equation}
The probability density~(\ref{eq:stationary-n4}) is the symmetric bimodal distribution with modes at $x=\pm \sigma^{\nicefrac{1}{3}}/\sqrt{2}$ and
the power-law asymptotics $P(|x|) \propto |x|^{-4}$.
The observed bimodality (\ref{eq:stationary-n4}) is related to the general property of the  L\'evy noise --- induced bifurcation in modality of the corresponding PDF for $t \rightarrow\infty$, see Refs.~\onlinecite{chechkin2002,chechkin2003,dubkov2007}.
In more general single-well potentials --- \bdt{NESS} (their PDFs) can be characterized by more than two modes  \cite{capala2019multimodal}.

The multimodality of \bdt{NESS} in overdamped systems calls to inquire whether PDFs in underdamped regime can be multimodal.
As it was shown in earlier works \cite{sokolov2011,zozor2011spectral}, for $V(x)=x^2/2$, \bdt{NESS} $P(x,v)$ are given by the 2D $\alpha$-stable density \cite{samorodnitsky1994,teuerle2012}, whose marginal \bdt{densities are  unimodal and given by} 1D $\alpha$-stable densities  --- in an analogy to their Gaussian white noise-driven cases.
\bdt{
Contrary to the stationary states in Gaussian white noise driven systems, see Eq.~(\ref{eq:bg}), under action of L\'evy noises two dimensional non-equilibrium stationary densities $P(x,v)$ do not factorize.
}
\bdt{
Therefore, position and velocity are not statistically independent \cite{sokolov2011}.
In Ref.~\onlinecite{sokolov2011}, due to divergence of covariance, the level of dependence was measured by the codifference \cite{samorodnitsky1994,wylomanska2015codifference}.
The nonlinear friction could increase the statistical dependence between position and velocity, as already the combined action of nonlinear friction and Gaussian white noise \cite{dechant2016heavytailed} introduces dependence. 
Moreover, we expect that in systems with full dynamics, analogously like in overdamped models \cite{garbaczewski2011levy,kusmierz2016breaking}, the condition of detailed balance is violated. 
Nevertheless, these issues (level of dependence and detailed balance) need further verification.}
In Ref.~\onlinecite{capala-underdamped}, we have extended studies \bdt{on underdamped systems under action of L\'evy noises} and have analyzed properties of \bdt{non-equilibrium} stationary PDFs for anharmonic potentials \bdt{in the case of linear damping} .
We have shown that in the system described by Eq.~(\ref{eq:full-langevin}), \bdt{i.e., in the regime of linear friction, the non-equilibrium} stationary state can be multimodal under the condition that damping is strong enough.
The constraint of the strong damping is related to the fact that for infinite damping ($\gamma\to\infty$) the motion described by Eq.~(\ref{eq:full-langevin}) becomes overdamped and 
 the corresponding L\'evy noise driven motion in single-well potentials (steeper than parabolic) \bdt{NESS} become at least ``bimodal'' for long times \cite{capala2019multimodal}.
In practical realizations though, this bimodality is observed for the finite damping.

The problem of multimodality of \bdt{NESS}, which is posed here, is related to the more general issue of existence of \bdt{NESS}.
We can ask the question what is the minimal steepness $n$ of the potential allowing for bounding of underdamped L\'evy flights.
The problem of the potential steepness is related to the friction.
The friction in Eq.~(\ref{eq:full-langevin}) is linear.
Consequently, the velocity $v$ changes according to
\begin{equation}
    \dot{v}(t)=-\gamma v(t) - V'(x)+ \zeta(t).
    \label{eq:velocity}
\end{equation}
\bdt{In general, steady state for the system described by Eq.~(\ref{eq:velocity}) is unknown.
Nevertheless, some intuitive insight might be gain by considering the motion of a free particle, i.e., the case where deterministic force $-V'(x)$ is omitted.}
If we disregard the deterministic force $-V'(x)$ in Eq.~(\ref{eq:velocity}) we get \bdt{the following equation}
\begin{equation}
\dot{v}(t)=-\gamma v(t) +\zeta(t),    
\label{eq:disregarded}
\end{equation}
\bdt{
which can be used to approximate $P(v)$ densities.
The quality of such approximation increases with the increase in $\gamma$, see Figs.~\ref{fig:X4V4g1} -- \ref{fig:X4V4g6}.
}
\bdt{Under such an approximation}, the evolution of the velocity $v(t)$ is described by the same equation like the evolution of the position $x(t)$ in the overdamped dynamics in the parabolic potential, see Eq.~(\ref{eq:langevin}).
Therefore, the \bdt{steady state} density $P(v)$ is given by the $\alpha$-stable density with the same stability index $\alpha$ like the noise $\zeta(t)$, see Refs.~\onlinecite{chechkin2002,chechkin2003,dybiec2007d}, and the rescaled scale parameter
\begin{equation}
\sigma = \frac{\sigma_0}{(\gamma \alpha)^{1/\alpha}},    
\label{eq:modsigma}
\end{equation}
where $\sigma_0$ is the scale parameter of the L\'evy noise $\zeta(t)$ in Eq.~(\ref{eq:disregarded}).
For $\gamma=0$ there is no stationary velocity distribution, but the velocity is still distributed according to the $\alpha$-stable density with the scale parameter growing in time as $\sigma(t)=\sigma_0 t^{\nicefrac{1}{\alpha}}$.
Consequently, for $\gamma=0$, there is no stationary state for the underdamped model described by Eq.~(\ref{eq:velocity}).
For $\gamma>0$, with the linear friction, the $P(v)$ density is \bdt{very well approximated} by the $\alpha$-stable density. Therefore, for the linear friction, the problem of existence of \bdt{NESS} for the model described by Eq.~(\ref{eq:full-langevin}) is equivalent to the problem of existence of \bdt{NESS} for the overdamped motion in $V(x)$, see Eq.~(\ref{eq:langevin}) and Ref.~\onlinecite{dybiec2010d}.
Consequently, for $n>2-\alpha$ \bdt{non-equilibrium stationary states} exist.

In more elaborate situation the friction term $T(v)$ in Eq.~(\ref{eq:full-langevin}) does not need to be linear \cite{oden1983nonlocal,pollak1993fokker,urbakh2004nonlinear,romanczuk2012active,lindner2007diffusion,lindner2010diffusion}. In such a case the Langevin equation (\ref{eq:full-langevin}) generalizes to
\begin{equation}
\left\{
\begin{array}{lcl}
    \dot{x}(t) & = & v(t) \\
    \dot{v}(t) & = & T(v) - V'(x)+ \zeta(t)
\end{array}
\right..
\label{eq:nonlinfricgen}
\end{equation}
As an example, the dynamical behavior of a mechanical system with dry friction has been described \cite{Elmer1997} by
\begin{equation}
T(v) =  -\gamma \sign(v)|v|^{\kappa-1}\;\;\;(\kappa>0)    .
\end{equation}
The linear friction corresponds to $\kappa=2$.
The friction $T(v)$ can be seen as an analog of the deterministic force $-V'(x)$ in the overdamped regime, compare Eq.~(\ref{eq:langevin}) and the second line of Eq.~(\ref{eq:nonlinfricgen}).
Consequently, it is possible to find the generalized $v$-potential.
Following this analogy, it is possible to relate the problem of existence of the \bdt{steady state} density $P(v)$  to the problem of existence of \bdt{NESS} in the overdamped dynamics.
Therefore, in order to bound velocity, the condition on $\kappa$ is the same as the condition on $n$ in $V(x)=|x|^n/n$, i.e.,
\begin{equation}
    \kappa > 2 -\alpha.
\end{equation}
Furthermore, for $\kappa > 4 -\alpha$, marginal densities $P(v)$ are characterized by the finite variance, see Refs.~\onlinecite{chechkin2002,dybiec2010d}.
Therefore, we can consider the sub-linear friction with $\kappa$ bounded from below, i.e., $2-\alpha<\kappa<2$.
In such a case the density $P(v)$  exists and most likely, for $n>2-\alpha$, a \bdt{non-equilibrium stationary state} $P(x,v)$ also exists.
The regime of super-linear friction, $\kappa>2$, is more transparent than the sub-linear case.
For $\kappa>2$, the density $P(v)$  asymptotically behaves as a power-law with lighter tails than noise in Eq.~(\ref{eq:nonlinfricgen}).
In other words, for $\kappa>2$, tails of $P(v)$ distribution decay faster than tails of the $\alpha$-stable density associated with the L\'evy noise $\zeta(t)$ in Eq.~(\ref{eq:full-langevin}).
For example, for $T(v)=-\gamma v^3$ with $\alpha=1$, asymptotics of $P(v)$ is $P(|v|) \propto |v|^{-4}$, see Eq.~(\ref{eq:stationary-n4}).
Therefore, we can speculate that the minimal exponent in the potential $V(x)=|x|^n/2$ is still bounded from below ($n>0$) but now it can be smaller than $2-\alpha$.
For instance, for $\kappa=4$ the variance of $P(v)$ is finite, therefore we expect that $P(x)$ exists for any $n>0$, what is confirmed by numerical simulations (results not shown).

In the next section (Sec.~\ref{sec:results}) we present results of our analysis of \bdt{non-equilibrium stationary states (NESS)} for anharmonic stochastic oscillators under nonlinear friction.
The manuscript is closed with Summary and Conclusions (Sec.~\ref{sec:summary}).

\section{Results\label{sec:results}}

\begin{figure}[!h]
\begin{center}
\includegraphics[width=0.75\columnwidth]{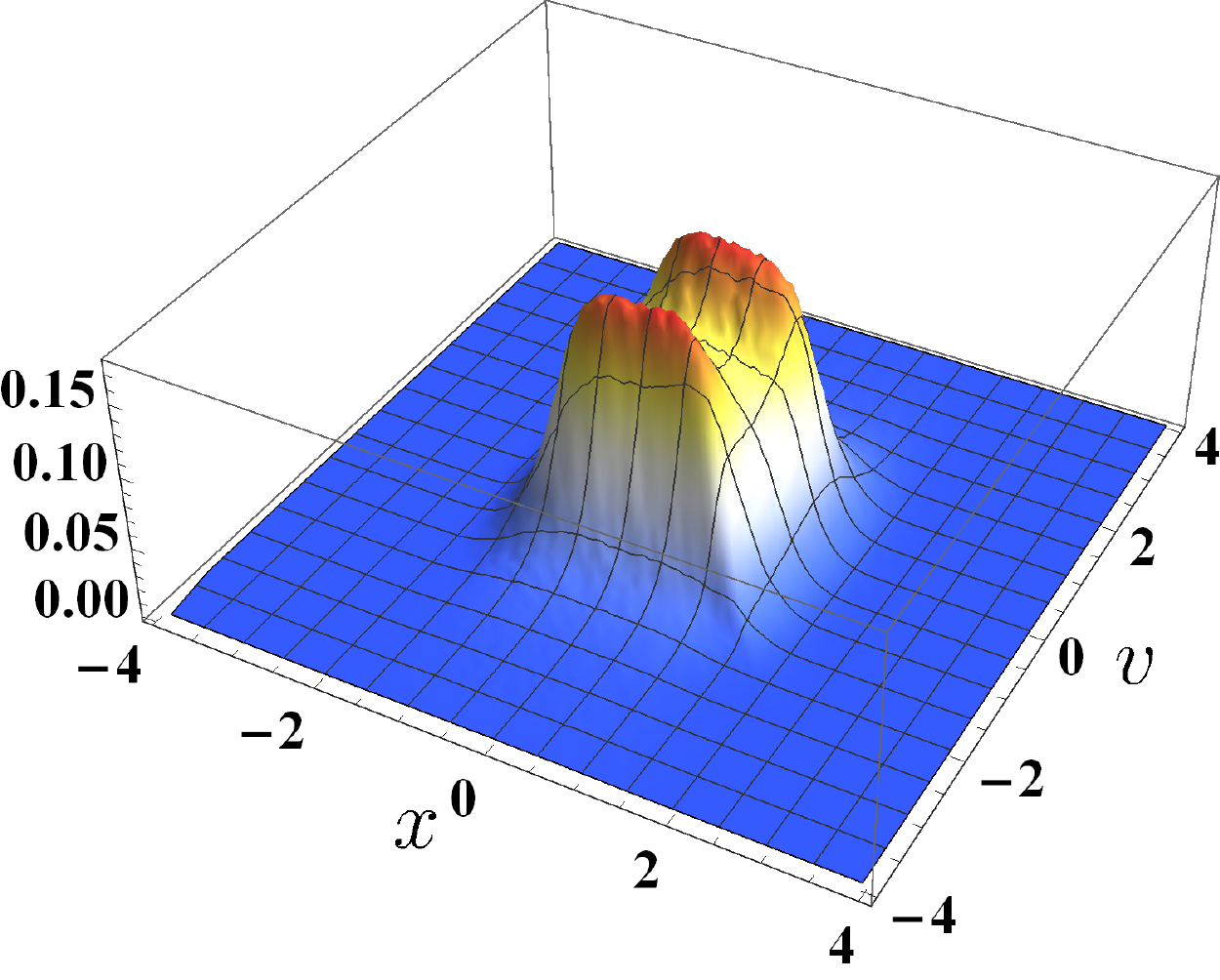} \\ \includegraphics[width=0.75\columnwidth]{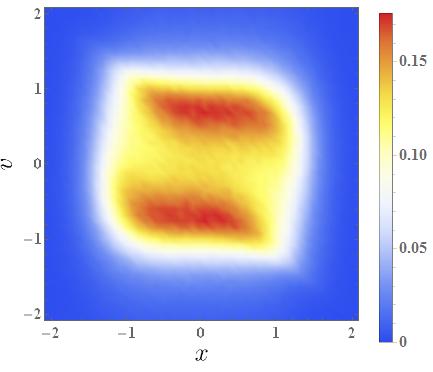}\\
\includegraphics[angle=0, width=0.3\textwidth]{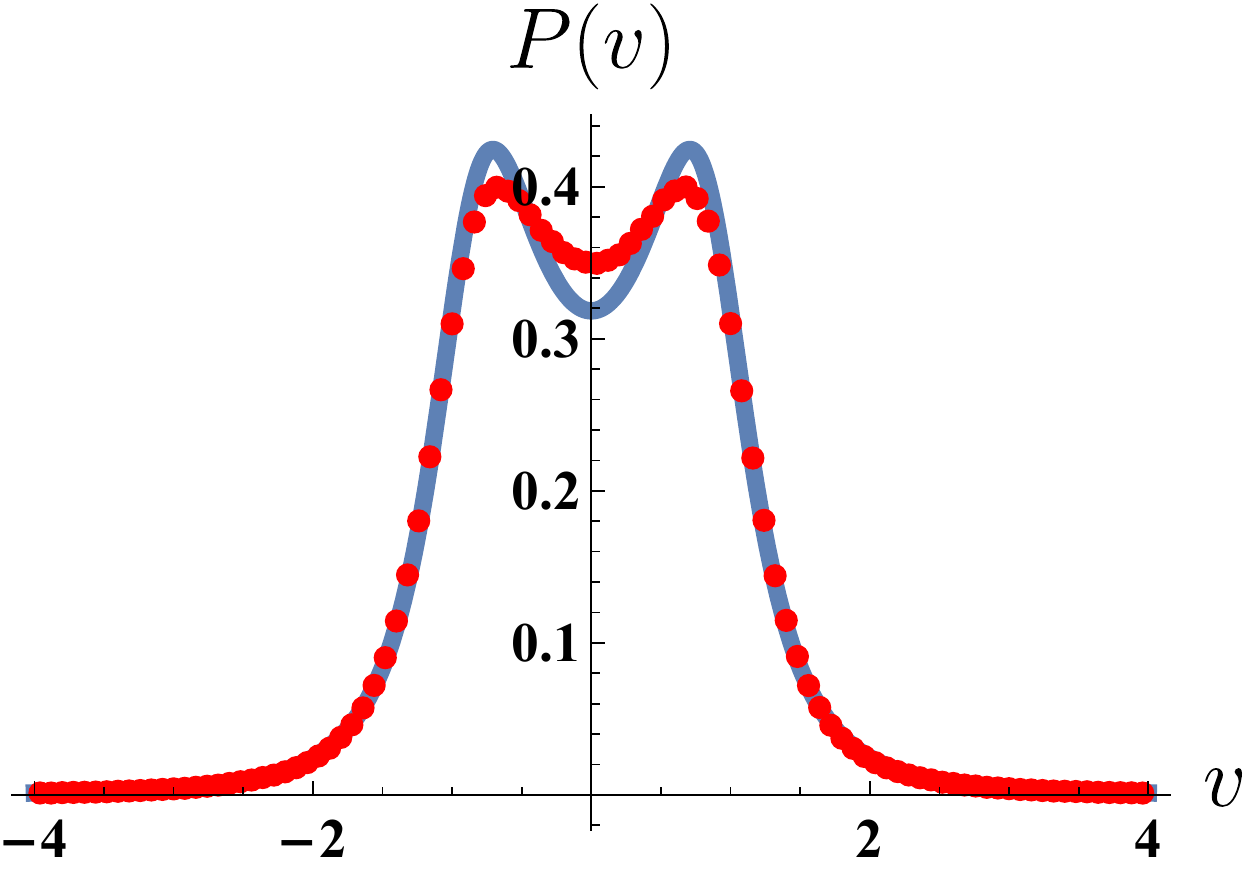} \\ \includegraphics[width=0.75\columnwidth]{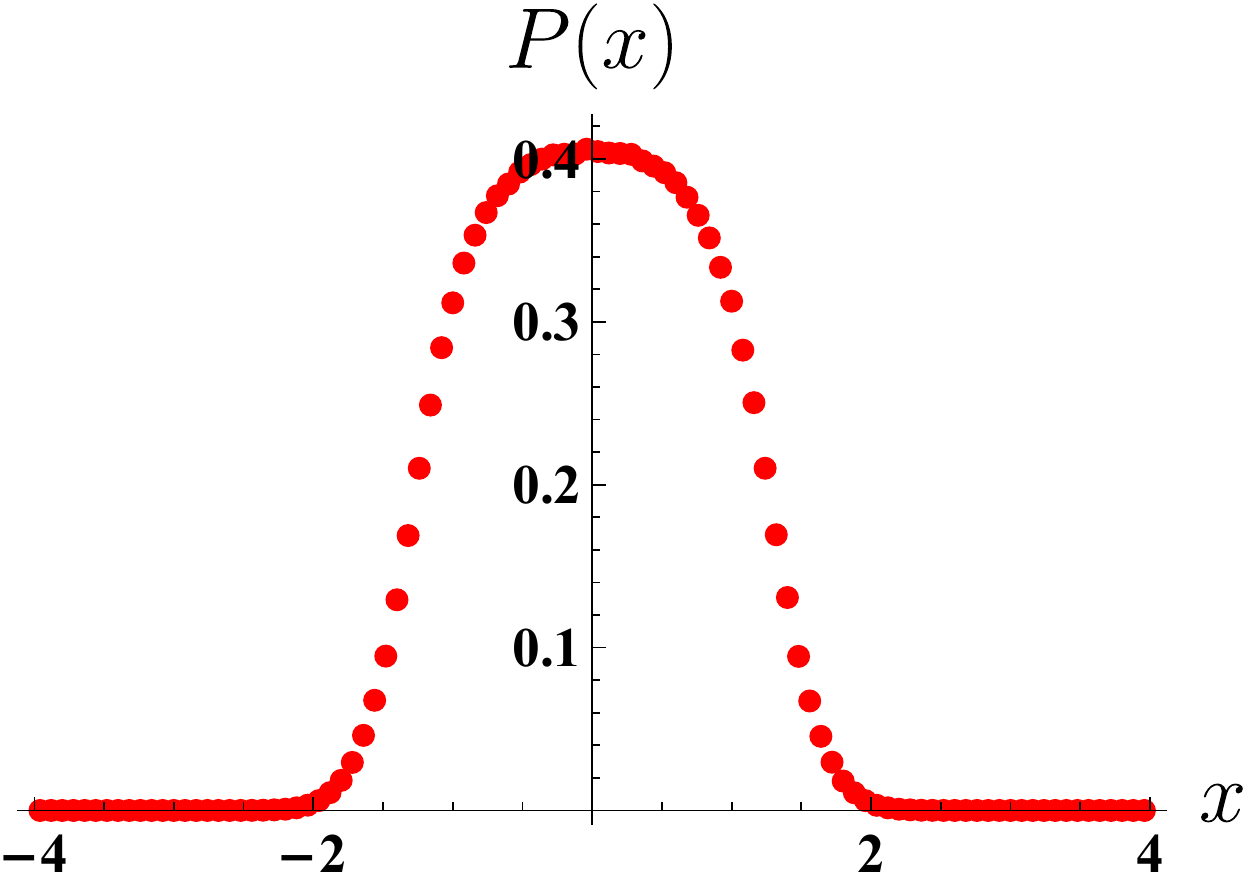}
 \end{center}
 \caption{\bdt{Non-equilibrium} stationary probability density $P(x,v)$ as the 3D-plot and the 2D  map (top panels), the velocity \bdt{non-equilibrium} stationary marginal density $P(v)$ (points) with the analytical solution (\ref{eq:stationary-n4}) with $\sigma$ given by Eq.~(\ref{eq:modsigma}) (solid line) and the position \bdt{non-equilibrium} stationary marginal density $P(x)$ (bottom panels).
 \bdt{The driving noise is the Cauchy noise, i.e., the L\'evy noise with $\alpha=1$.}
 The damping parameter $\gamma$ is set to $\gamma=1$.
 }
 \label{fig:X4V4g1}
\end{figure}

\begin{figure}[!h]
\begin{center}
\includegraphics[width=0.75\columnwidth]{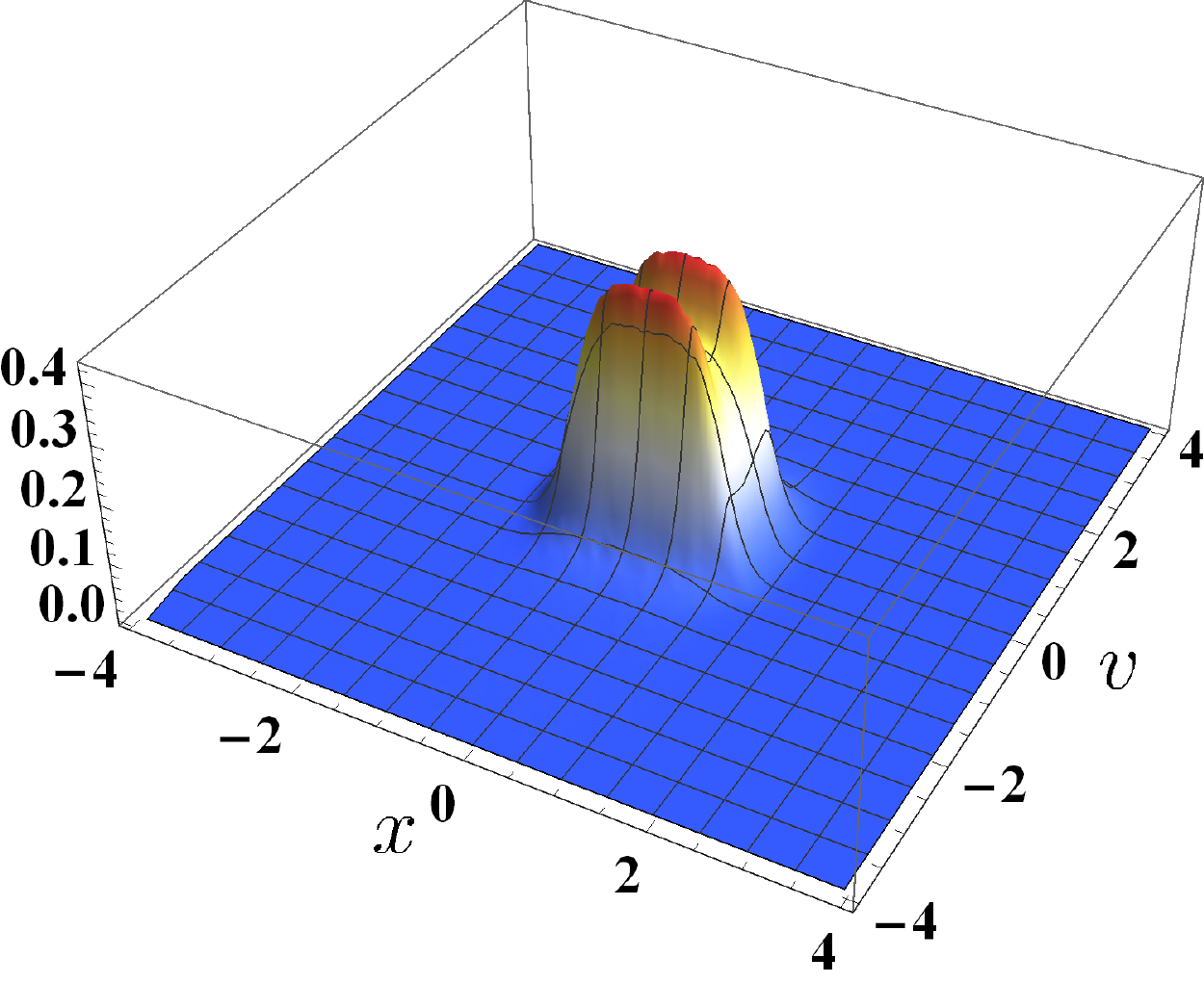} \\ \includegraphics[width=0.75\columnwidth]{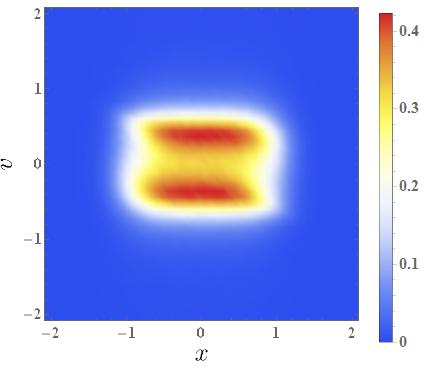}\\
\includegraphics[width=0.75\columnwidth]{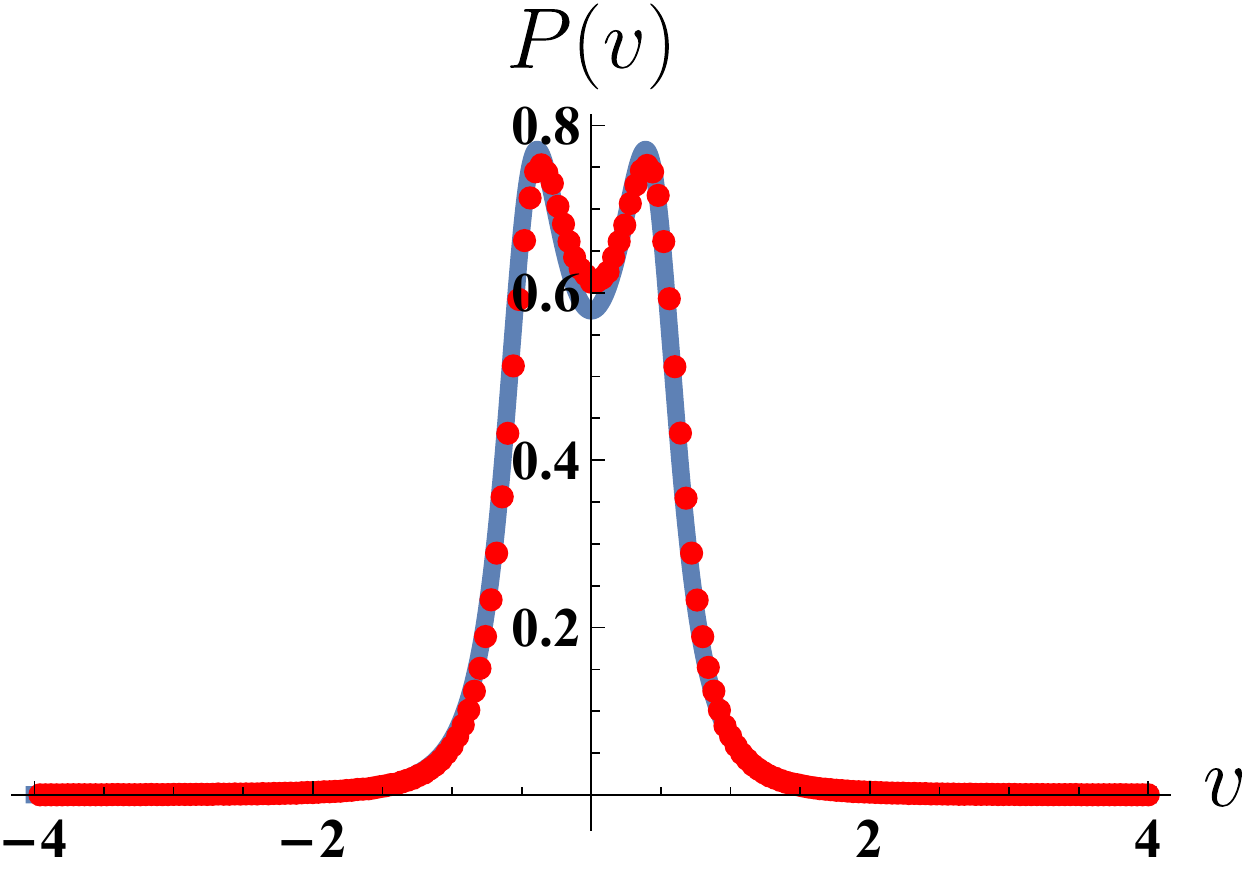} \\ \includegraphics[width=0.75\columnwidth]{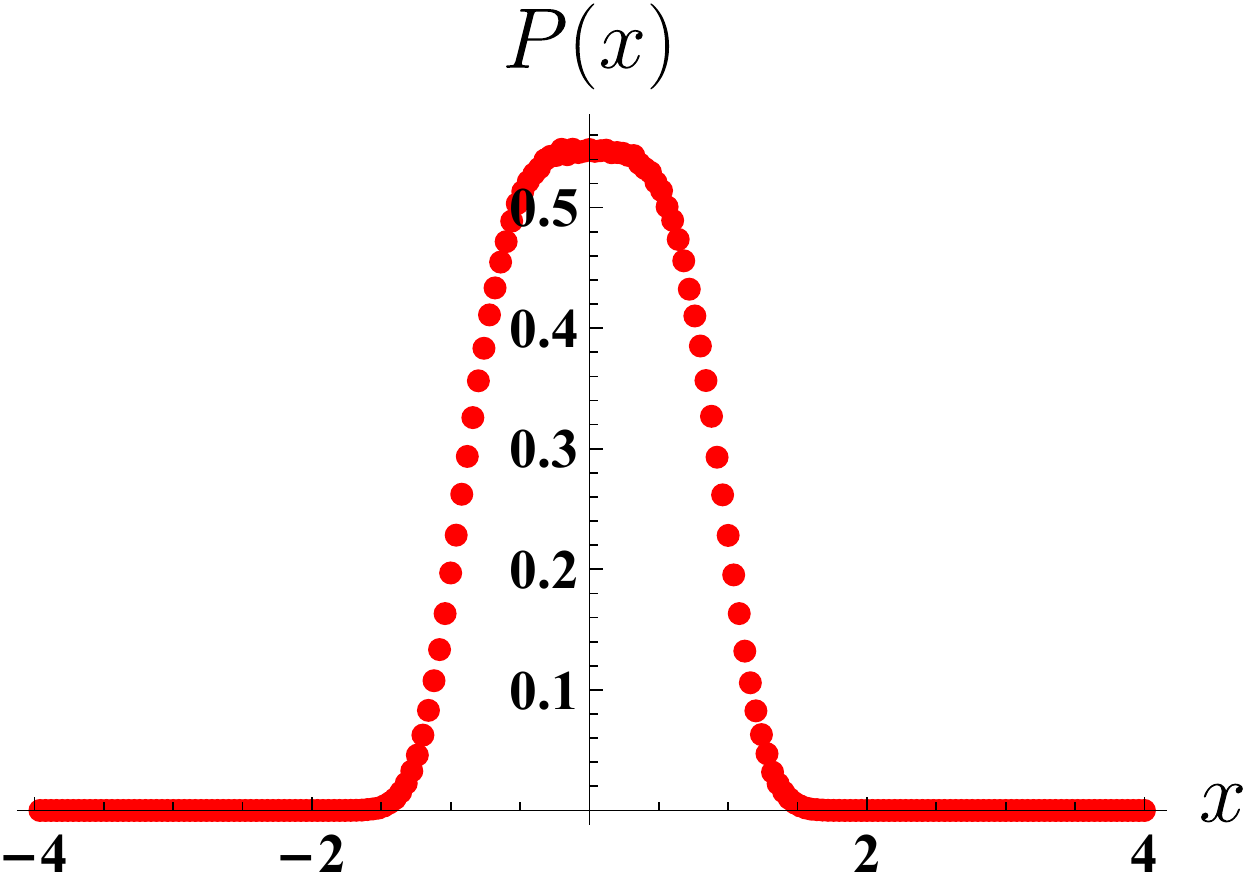}
 \end{center}
 \caption{The same as in  Fig.~\ref{fig:X4V4g1} for $\gamma=6$.
 }
 \label{fig:X4V4g6}
\end{figure}

In what follows, we relax the assumption of linear friction and assume that friction depends nonlinearly on the particle velocity.
We start with $T(v)=-\gamma v^3$.
Such a system is described by the following Langevin equation
\begin{equation}
\left\{
\begin{array}{lcl}
    \dot{x}(t) & = & v(t) \\
    \dot{v}(t) & = & -\gamma v^3(t) - x^3(t)+ \zeta(t) \\
\end{array}
\right..
\label{eq:nonlinfric}
\end{equation}
Results of simulations depicted in Figs.~\ref{fig:X4V4g1} and~\ref{fig:X4V4g6} are significantly different from results for the linear friction with the same potential $V(x)$ \bdt{and the same noise, i.e., the Cauchy noise ($\alpha=1$)}, see Ref.~\onlinecite{capala-underdamped}.
\bdt{If one disregard} the deterministic $-x^3$ force, the Langevin equation for the velocity evolution becomes similar to the overdamped equation~(\ref{eq:langevin_n4}) with the position $x$ replaced by the velocity $v$.
Therefore, \bdt{we could expect that}, analogously to the \bdt{bimodal non-equilibrium} stationary density $P(x)$ associated with Eq.~(\ref{eq:langevin_n4}), the velocity marginal density $P(v)$ becomes also bimodal. 
This bimodality is also reflected in the shape of the full probability density:
In the top panel of Figs.~\ref{fig:X4V4g1} and~\ref{fig:X4V4g6}, there are two maxima separated only in the velocity direction. 
For $\gamma=1$ (Fig.~\ref{fig:X4V4g1}), there is no multimodality in the position \bdt{marginal} PDF. 
Contrary to the case of linear friction, see Ref.~\onlinecite{capala-underdamped}, the increase in  $\gamma$ does not induce bimodal \bdt{steady} states in the position marginal distribution even for $\gamma=6$ (Fig.~\ref{fig:X4V4g6}). 
The change in $\gamma$ affects only widths of marginal distributions but it does not change its modality tails' asymptotics. 
When the damping increases, position probability densities $P(x)$ become localized around minimum of the deterministic potential.

\begin{figure}[!h]
\begin{center}
\includegraphics[width=0.75\columnwidth]{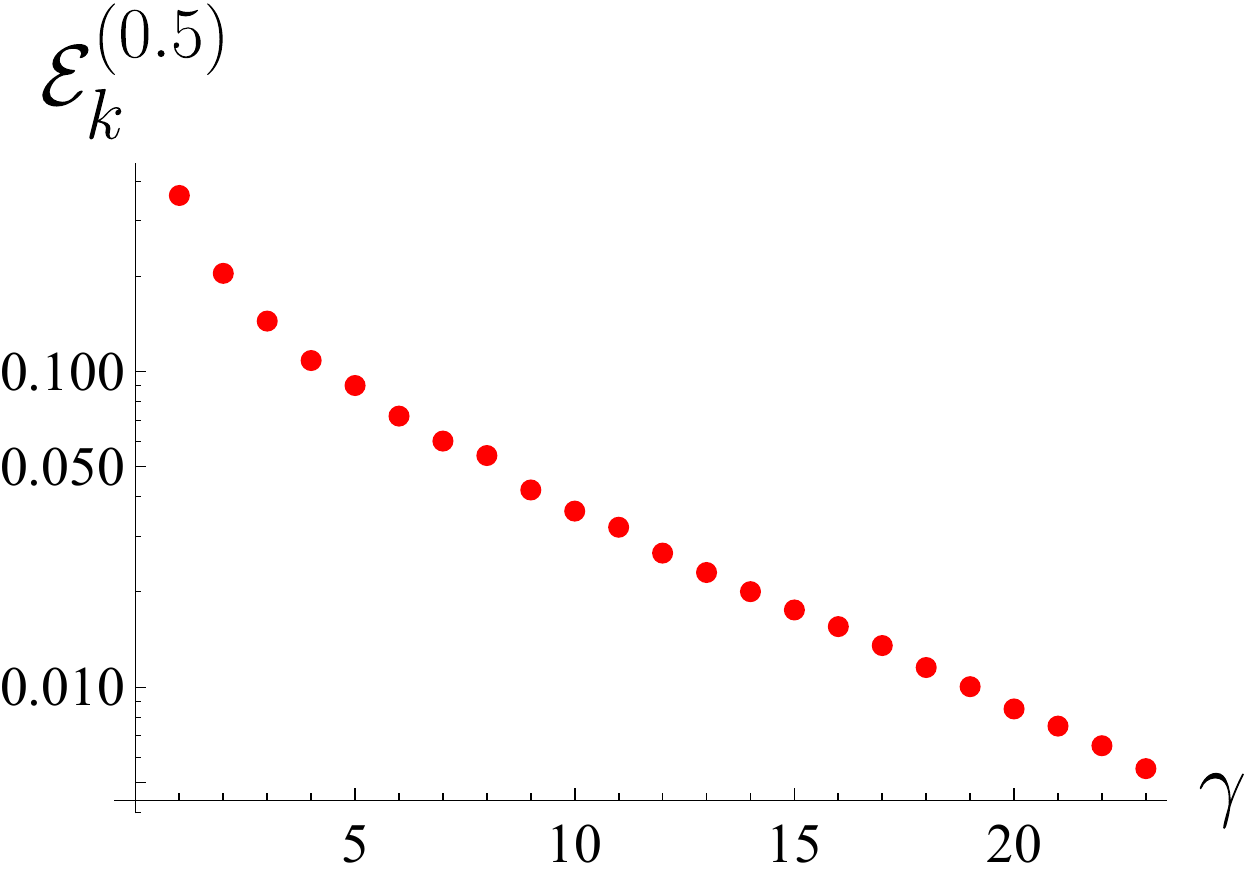} \\ \includegraphics[width=0.75\columnwidth]{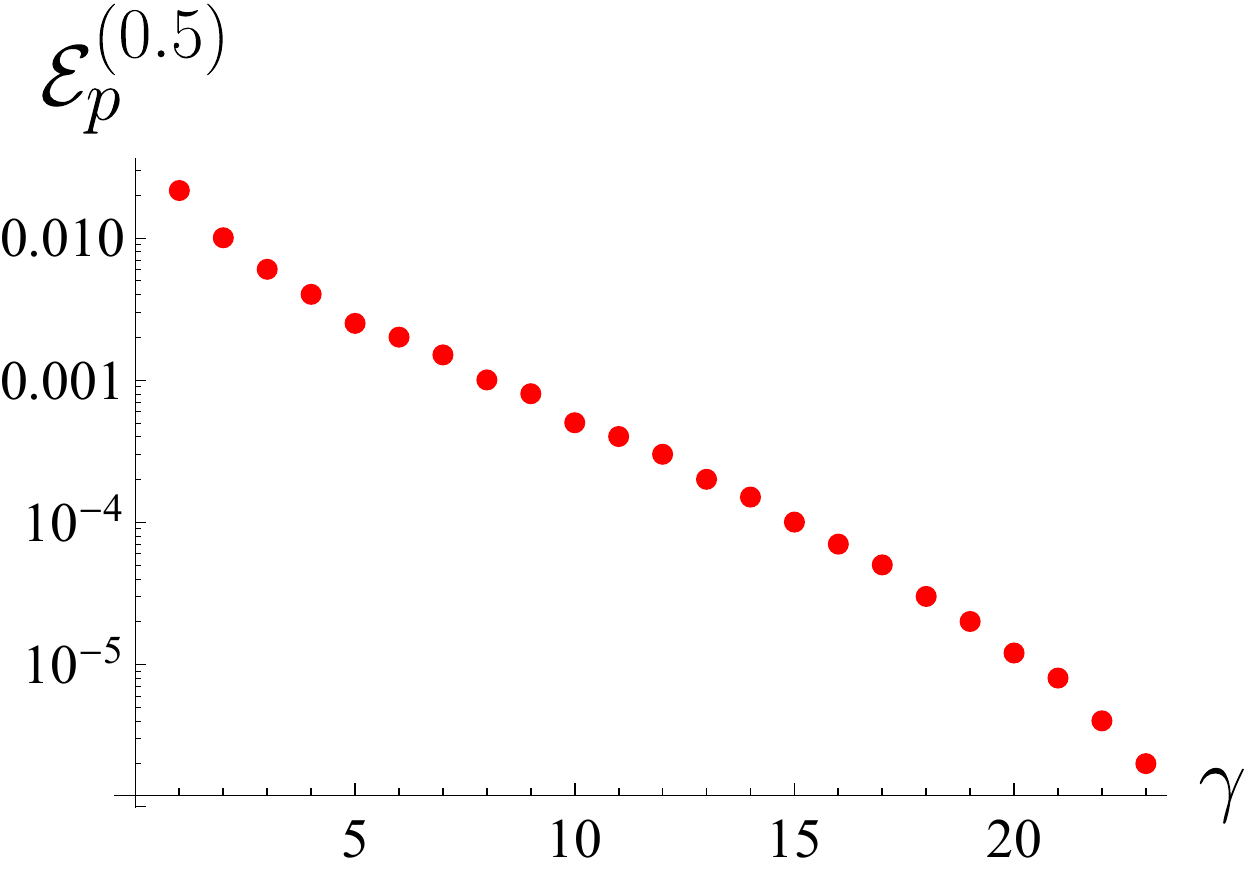}
 \end{center}
 \caption{Median of the kinetic energy $\mathcal{E}_k^{(0.5)}$ (top panel) and potential energy $\mathcal{E}_p^{(0.5)}$
 (bottom panel) as a function of friction parameter $\gamma$ for the system described by Eq.~\eqref{eq:nonlinfric}.
    }
 \label{fig:Energia_X4V4g1}
\end{figure}

\begin{figure}[!h]
\begin{center}
\includegraphics[width=0.98\columnwidth]{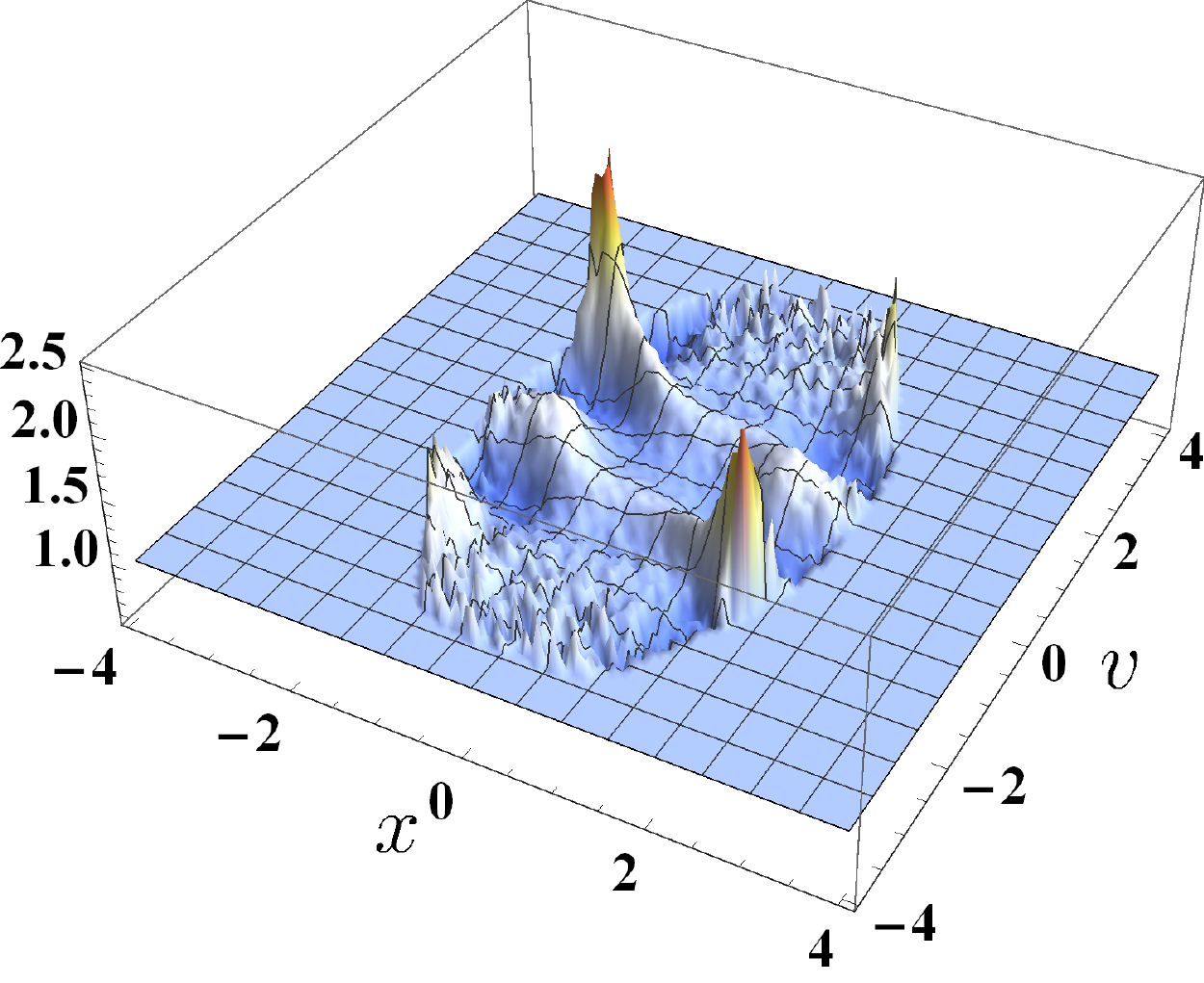} \\
 \end{center}
 \caption{
 \bdt{Ratio $P(x,v)/[P(x)P(v)]$, see Fig.~\ref{fig:X4V4g1}, quantifying departure from the corresponding Boltzmann-Gibbs equilibrium stationary states, for which $P(x,v)/[P(x)P(v)]\equiv 1$.}
    }
 \label{fig:ratioV3}
\end{figure}

\begin{figure}[!h]
\begin{center}
\includegraphics[width=0.75\columnwidth]{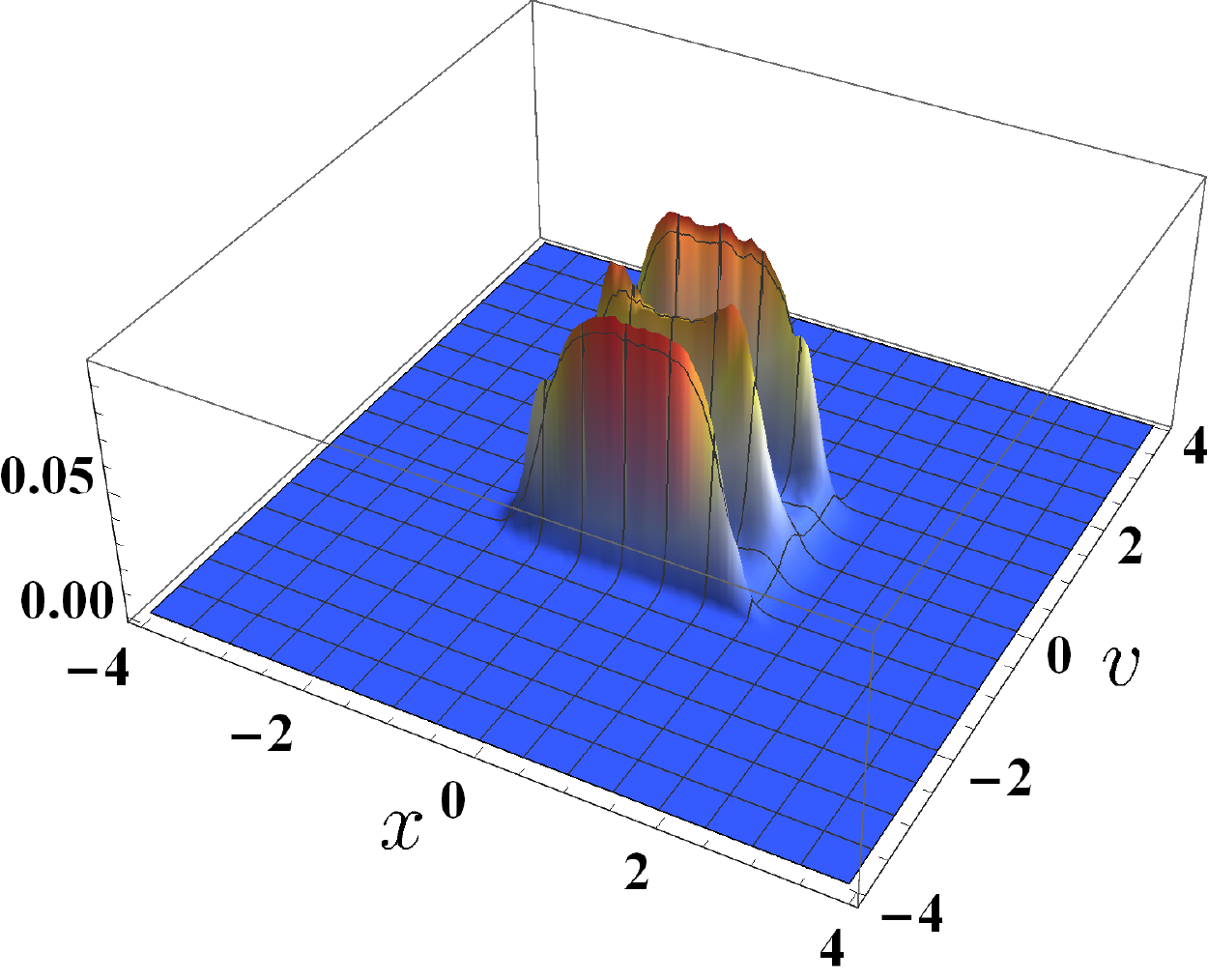} \\ \includegraphics[width=0.75\columnwidth]{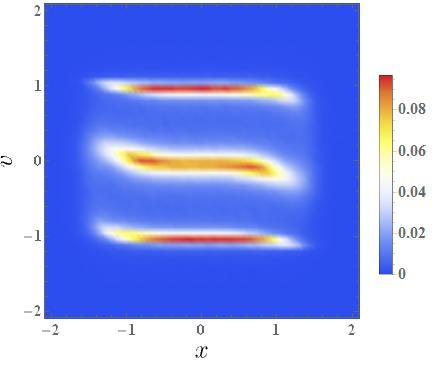}\\
\includegraphics[width=0.75\columnwidth]{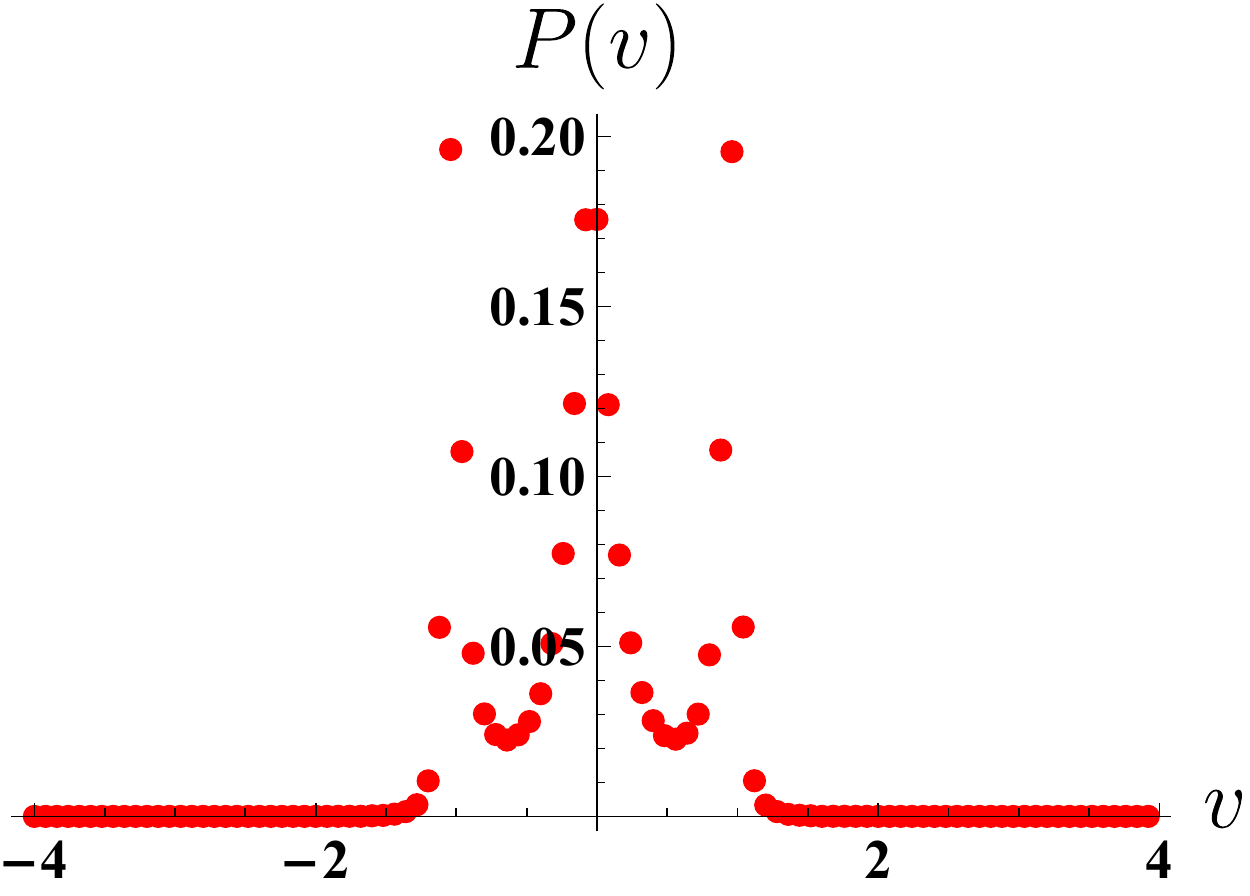} \\ \includegraphics[width=0.75\columnwidth]{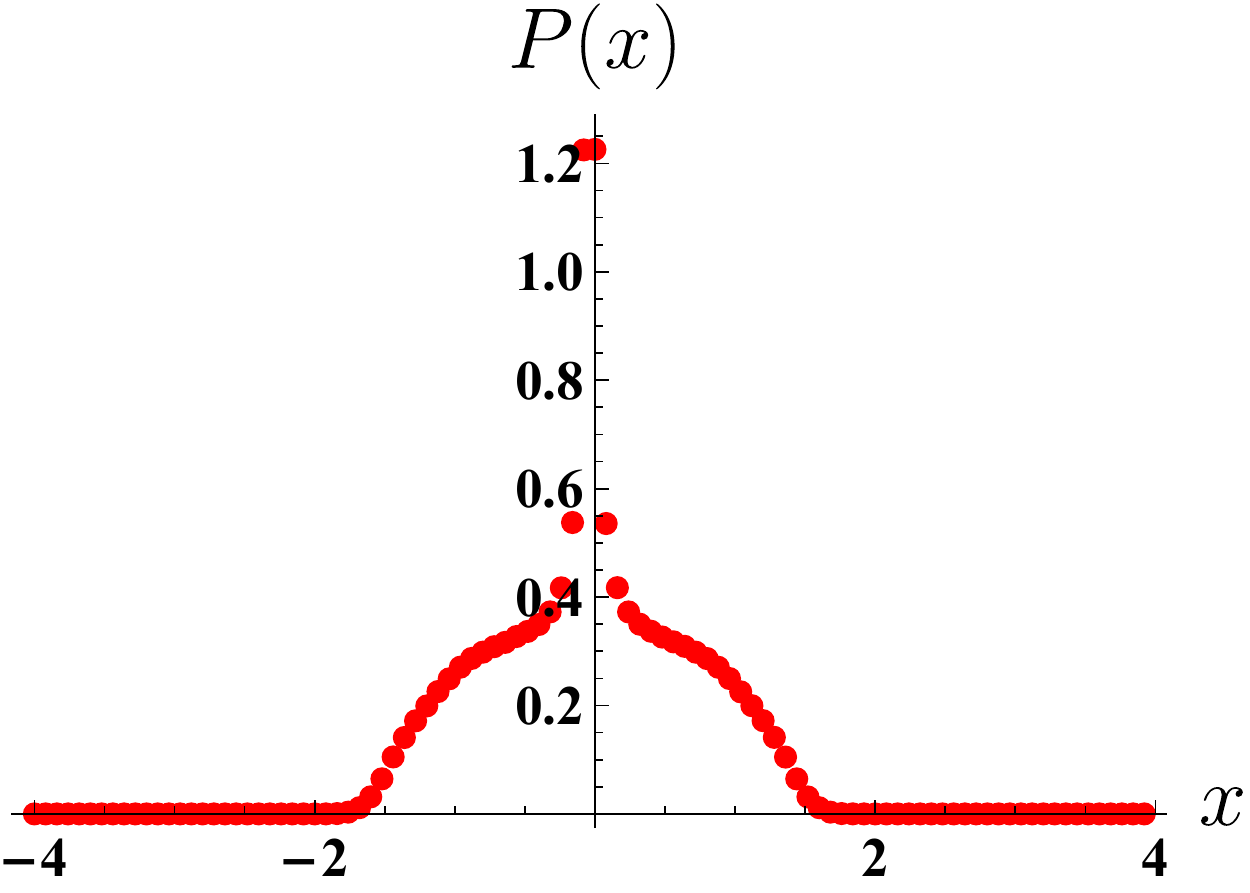}
 \end{center}
 \caption{The \bdt{non-equilibrium stationary states} and marginal densities for $T(v)$ given by  Eq.~(\ref{eq:nonmonotonousfriction}) with $\gamma=4$.
 \bdt{The driving noise is the Cauchy noise, i.e., the L\'evy noise with $\alpha=1$}.}
 \label{fig:X4V6g4}
\end{figure}

\bdt{Non-equilibrium stationary states (NESS)} in systems characterized by the nonlinear friction can be very different from similar models with the linear friction \cite{romanczuk2012active}.
\bdt{For instance, this can demonstrated} by taking the limit of strong damping in Eq.~(\ref{eq:nonlinfric}).
Assuming stationarity in the velocity, i.e., $\dot{v}(t)$=0 in Eq.~(\ref{eq:nonlinfric}), the Langevin equation reduces to
\begin{equation}
\gamma \dot{x}^3(t) = -x^3(t) + \zeta (t),
    \label{eq:nonlinfricod}
\end{equation}
which is significantly different from Eq.~(\ref{eq:langevin_n4}).
\bdt{In the regime of nonlinear  friction, the overdamped Langevin equation contains the nonlinear $\dot{x}^3(t)$ term, which is absent in the regime of linear friction, cf., Eq.~(\ref{eq:langevin_n4}) and Eq.~(\ref{eq:nonlinfricod}).}
Consequently, \bdt{not only equations but} also position marginal \bdt{non-equilibrium} stationary densities $P(x)$ are very different from their underdamped counterparts, see bottom panels of Figs.~\ref{fig:X4V4g1},~\ref{fig:X4V4g6} and especially of Fig.~\ref{fig:X4V6g4}.
\bdt{More precisely, $P(x)$  densities in Figs.~\ref{fig:X4V4g1} -- \ref{fig:X4V4g6} are unimodal, while for the same potential with the linear friction and under the action of the same noise, they are bimodal.
These differences can be easily tracked in the strong damping limits, see Eqs.~(\ref{eq:langevin_n4}) and~(\ref{eq:nonlinfricod}), and they are also well visible for finite $\gamma$ also, see Figs.~\ref{fig:X4V4g1} -- \ref{fig:X4V4g6}.
}

\begin{figure}[!h]
\begin{center}
\includegraphics[width=0.75\columnwidth]{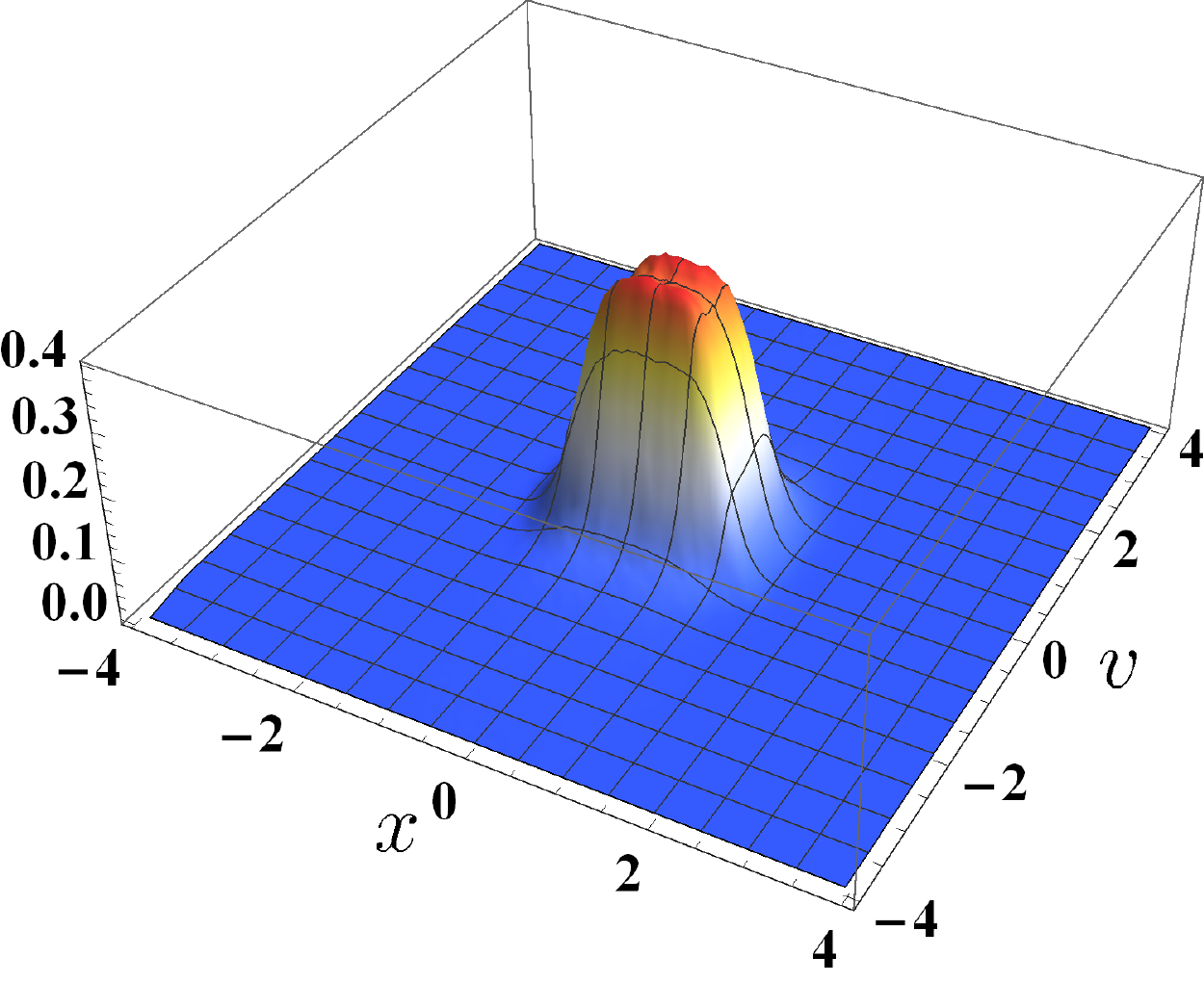} \\ \includegraphics[width=0.75\columnwidth]{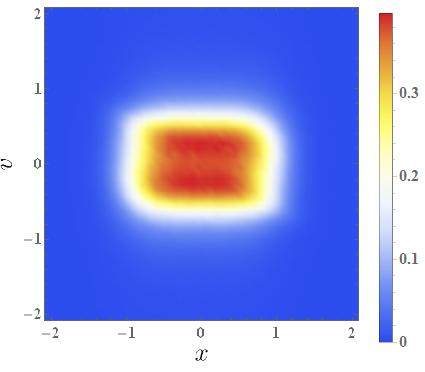}\\
\includegraphics[width=0.75\columnwidth]{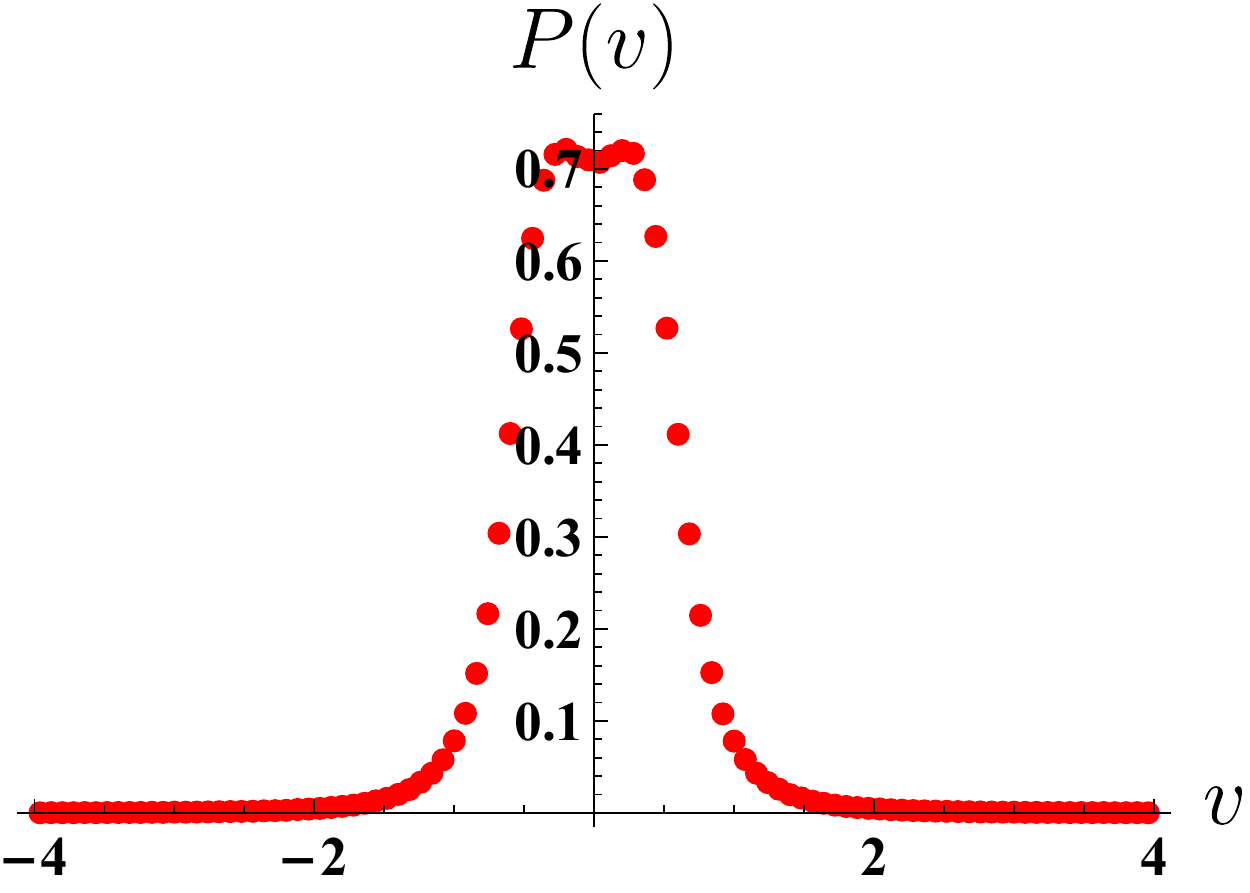} \\ \includegraphics[width=0.75\columnwidth]{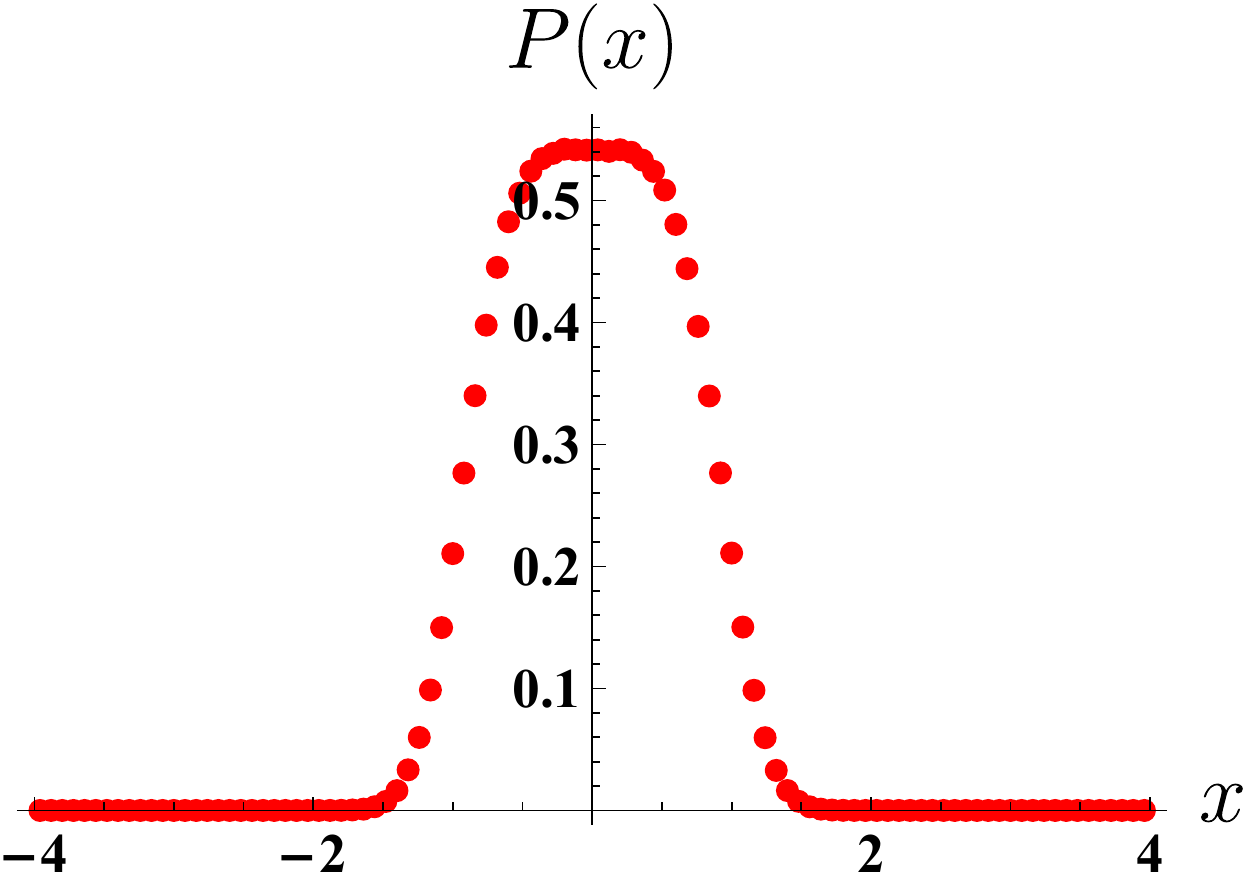}
 \end{center}
  \caption{\bdt{Non-equilibrium stationary probability density $P(x,v)$ as the 3D-plot and the 2D map (top panels), the velocity non-equilibrium stationary marginal density $P(v)$ and the position non-equilibrium stationary marginal density $P(x)$ (bottom panels).
 The driving noise is the Cauchy noise, i.e., the L\'evy noise with $\alpha=1$, while the friction term is given by Eq.~\eqref{eq:fric4p2} with $\gamma=4$ and $a=0.2$.
 }}
 \label{fig:X4V4g4b02}
\end{figure}

\begin{figure}[!h]
\begin{center}
\includegraphics[width=0.75\columnwidth]{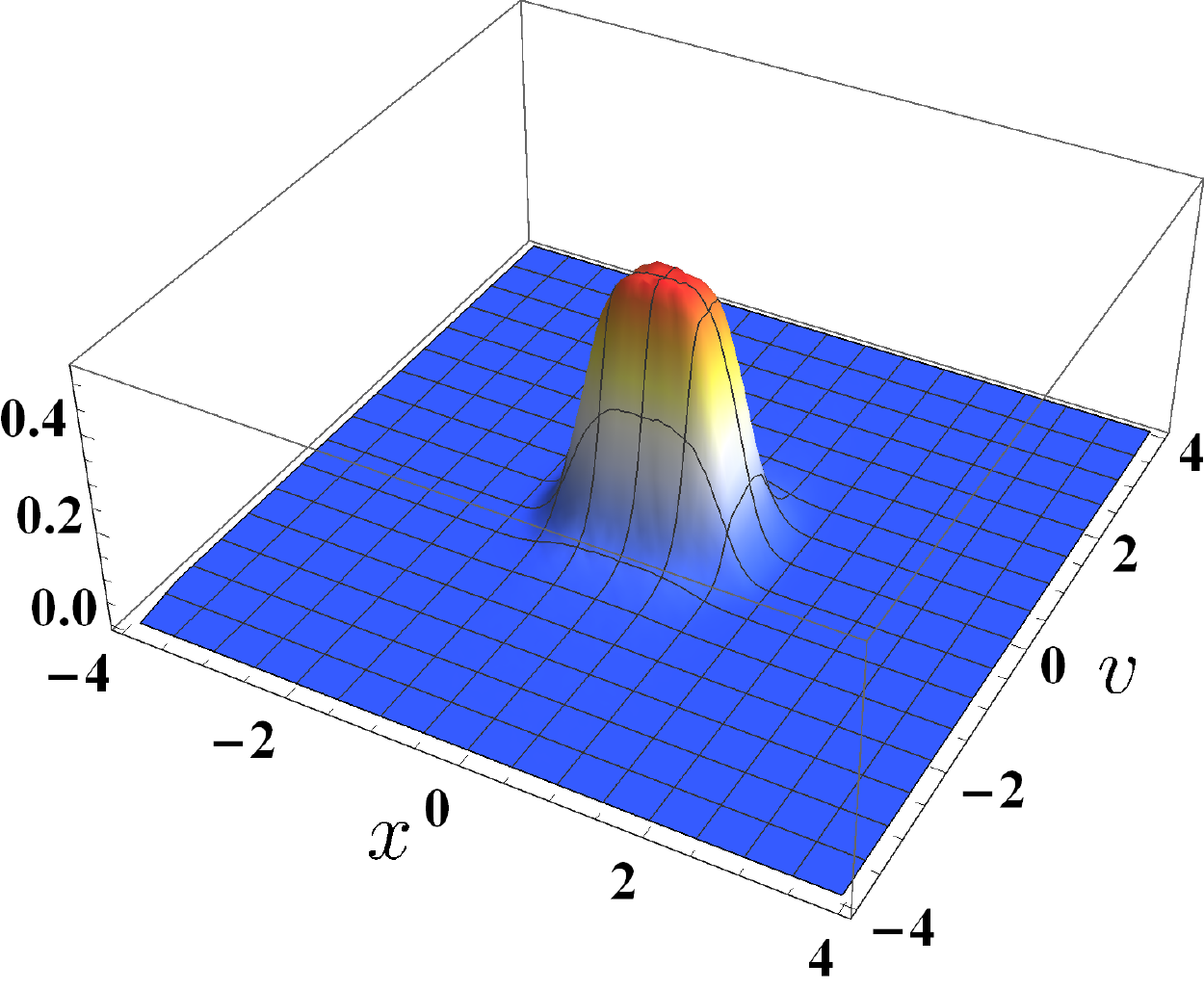} \\ \includegraphics[width=0.75\columnwidth]{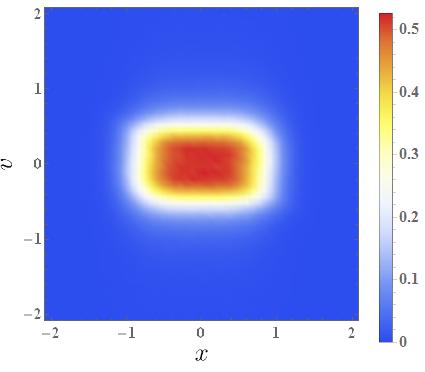}\\
\includegraphics[width=0.75\columnwidth]{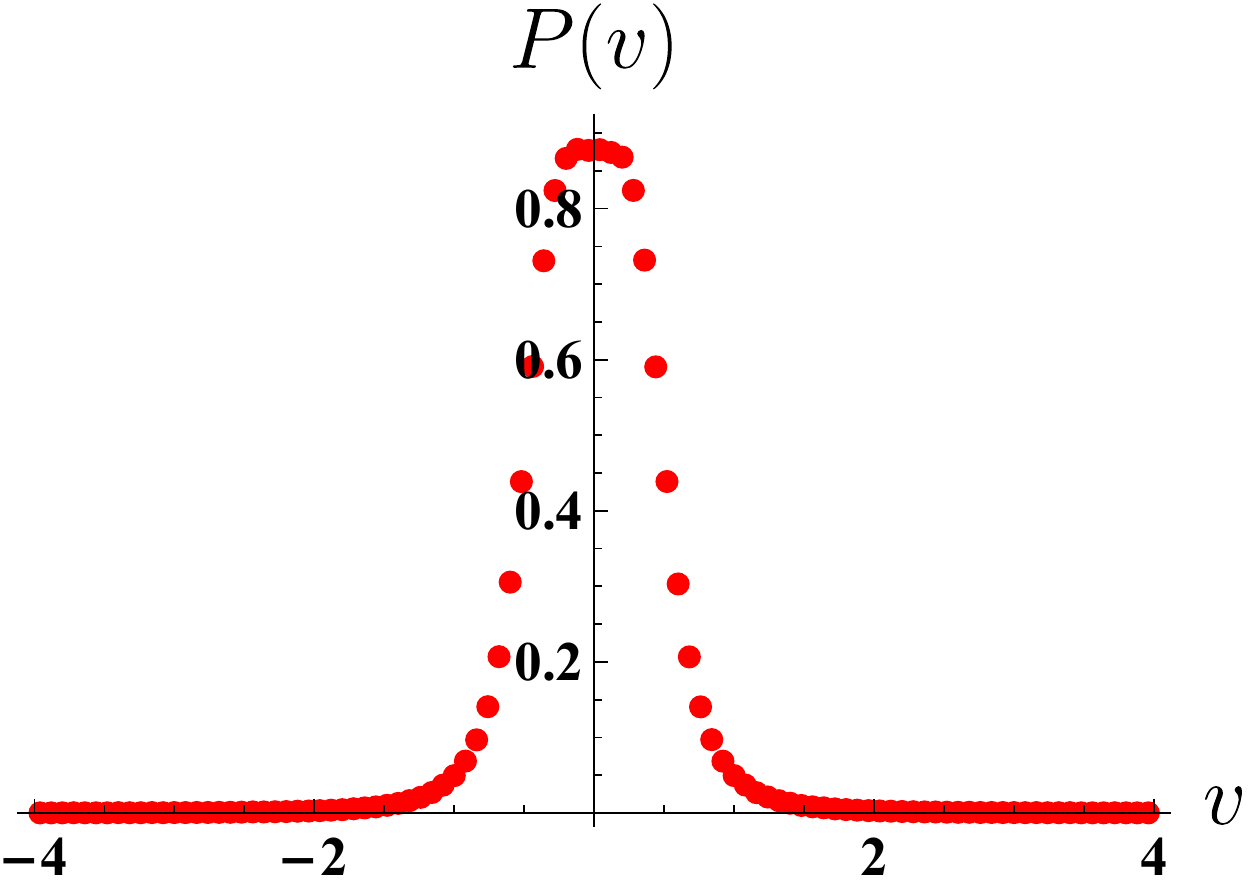} \\ \includegraphics[width=0.75\columnwidth]{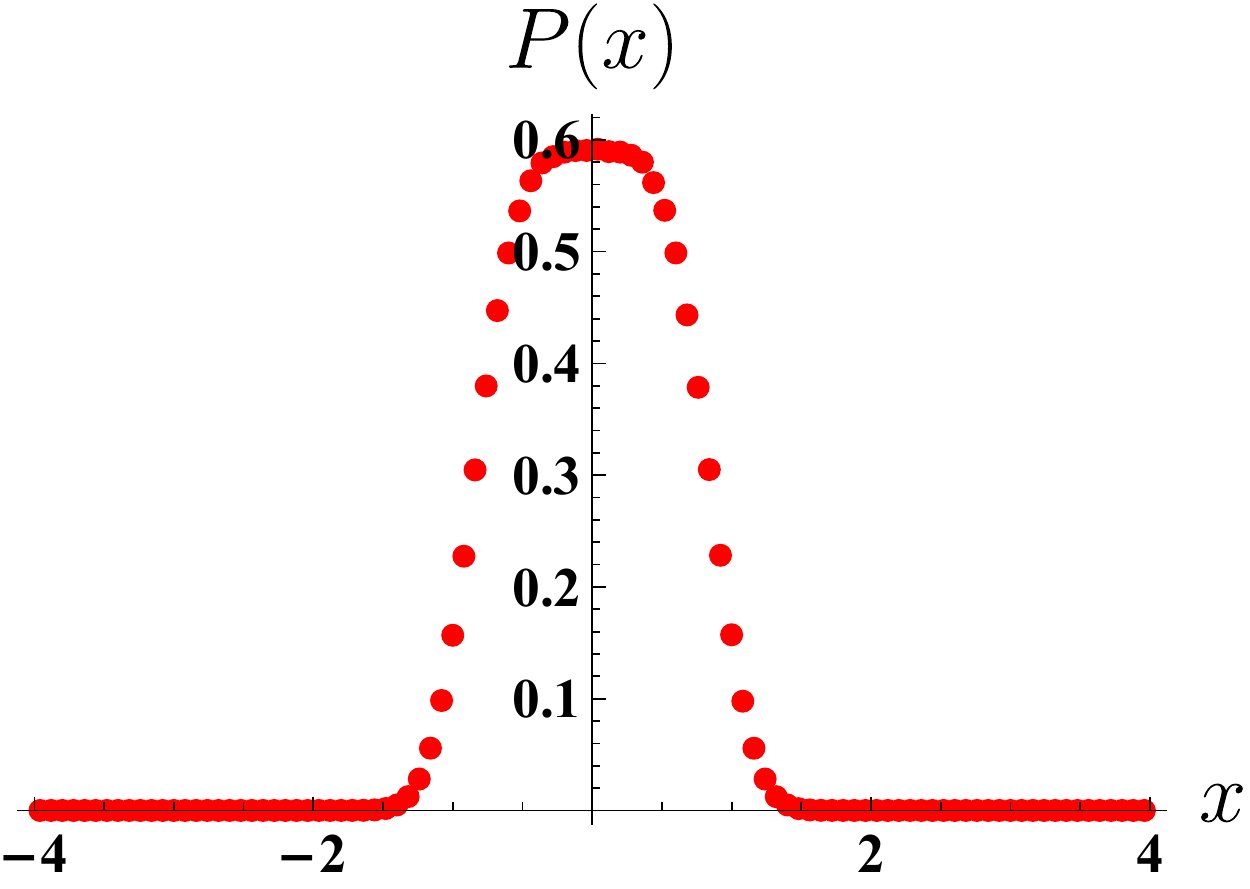}
 \end{center}
 \caption{The same as in  Fig.~\ref{fig:X4V4g4b02} for $\gamma=6$.}
 \label{fig:X4V4g6b02}
\end{figure}
\afterpage{\clearpage}

In analogy to the Langevin dynamics with linear friction, the lack of \bdt{bimodality} in the position marginal distribution can be better understood in terms of the analysis of the velocity marginal distribution:
Even for the large value of the friction parameter $\gamma$, the velocity distribution is bimodal. 
Therefore, occurrence of non-zero velocity is very likely and consequently more trajectories visit $x=0$ as large velocity helps to reach the origin.
Therefore, instead of minimum of $P(x)$ at the origin, there is a maximum.
In the deterministic dynamics, i.e., for $\zeta(t)\equiv 0$, the nonlinear damping secures observation of long lasting, persistent oscillations in $x(t)$, even at large values of $\gamma$.
As a result, and contrary to the overdamped case, a trajectory reaches the potential minimum in a finite time.
In consequence, the likelihood of returning to initial position before next ``long jump'' is not negligible.
Existence of two modal values in the velocity marginal distribution may be attributed to these oscillations which occur between noise pulses inducing transition between the modes \cite{dybiec2018conservative}.

The velocity marginal distributions $P(v)$ depicted in Figs.~\ref{fig:X4V4g1} and \ref{fig:X4V4g6} deviate from the analytical solution (\ref{eq:stationary-n4}) with the scale parameter given by Eq.~(\ref{eq:modsigma}), especially in the central part. 
Eq.~(\ref{eq:stationary-n4}) is the solution of the Eq.~(\ref{eq:disregarded}), which differs from the second line of Eq.~(\ref{eq:nonlinfric}) by disregarding the deterministic force, \bdt{while simulations are performed for the whole dynamics, i.e., the deterministic force $-V'(x)$ is also taken into account.}
The discrepancy between results of simulations and \bdt{non-equilibrium} stationary density given by Eq.~(\ref{eq:stationary-n4}) is produced by the deterministic force $-V'(x)$. In the force free case, $-V'(x) \equiv 0$, a perfect agreement is observed.
Moreover, with the increasing damping the level of disagreement decreases, see Fig.~\ref{fig:X4V4g1} and \ref{fig:X4V4g6},  because for larger $\gamma$ the velocity distribution is narrower and, most importantly, it equilibrates faster \cite{chechkin2000linear}.

Let us further analyze statistical properties of  kinetic $\mathcal{E}_k$ and potential $\mathcal{E}_p$ energies of such a  system.
Analogously to position $x$ and velocity $v$, also kinetic $\mathcal{E}_k$ and potential $\mathcal{E}_p$ energies are now random variables.
Their distributions can be calculated by use of PDFs $P(x,v)$ and suitable transformation of variables.
Due to a $|v|^{-4}$ asymptotic of the velocity marginal distribution, see Eq.~(\ref{eq:nonlinfric}), the mean value of kinetic energy ($\langle \mathcal{E}_k \rangle = \langle v^2 \rangle/2$) exists.
Moreover, a very fast decay of tails of the position PDF suggests that also the mean value of the potential energy ($\langle \mathcal{E} \rangle_p= \langle V(x) \rangle$) should exist.
As it is demonstrated in figures, velocity and positions distribution are well localized.
Nevertheless, it is very difficult to calculate numerically mean values of kinetic and potential energies.
Regardless of the  integration time step $\Delta t$, there is a non-negligible probability of observing very strong noise pulses which are responsible for the occurrence of very large velocities and long displacements resulting in the possibility of reaching distant positions.
These extreme events make the numerical calculation of the average energies ill posed.
Already a single extreme observation makes averages to explode in an uncontrollable way.
Therefore, instead of calculating averages, we have employed medians of kinetic ($\mathcal{E}_k^{(0.5)}$) and potential ($\mathcal{E}_p^{(0.5)}$) energies as they are robust parameters to rare but extreme events (outliers). 
Fig.~\ref{fig:Energia_X4V4g1}, \bdt{please note log-linear scale}, presents medians of energy distributions as  functions of damping parameter $\gamma$ for the process described by Eq.~\eqref{eq:nonlinfric}.
Both medians $\mathcal{E}_k^{(0.5)}$ and $\mathcal{E}_p^{(0.5)}$ \kc{exponentially} decrease with the increasing $\gamma$. 
Note that  the median of kinetic energy $\mathcal{E}_k^{(0.5)}$ is about order of magnitude larger then the median of potential energy $\mathcal{E}_p^{(0.5)}$.
\kc{One may also observe that $\mathcal{E}_p^{(0.5)}$ decays faster that $\mathcal{E}_k^{(0.5)}$.}
This difference may be deduced from marginal distributions: First, most of the probability mass is located in the $(-1,1)$ interval, both for position and velocity. 
Therefore, due to the relation between the velocity $v$ and the kinetic energy ($\mathcal{E}_k=v^2/2$) and the position $x$ and the potential energy ($\mathcal{E}_p=x^4/4$), most of the probability mass for energies is located in the $[0,1)$ intervals. 
For the argument from the $[0,1)$ interval, the function $x^4$ increases slower than $v^2$, thus, if the velocity and position marginal distribution were the same, one could expect the lower value of the median of the potential energy than the corresponding median of the kinetic energy.
However, both distributions differ significantly. 
Due to fast-decaying tails of the position marginal distribution, probability mass for the potential energy is concentrated near 0. 
At the same time, for the kinetic energy, the probability mass is moved towards larger values of $v$, because of power-law tails and bimodality of the velocity probability density.
These differences produce significantly higher value of the median of kinetic energy in comparison to the median of the potential energy. 

\bdt{
In Ref.~\onlinecite{sokolov2011} it was demonstrated that the non-equilibrium stationary states for the L\'evy harmonic oscillator (under linear friction) are given by the 2D $\alpha$-stable densities and position and velocity are not statistically independent.
The analogous situation is observed for anharmonic L\'evy oscillators in the regime of nonlinear friction, which are studied within the current manuscript.
In order to prove statistical dependence of $x$ and $v$ we have plotted the ratio of the full non-equilibrium stationary probability density $P(x,y)$ and marginal densities $P(x), P(v)$, i.e., $P(x,y)/[P(x)P(y)]$, see Fig.~\ref{fig:ratioV3}.
Close inspection of Fig.~\ref{fig:ratioV3} clearly indicates that position $x$ and velocity $v$ are not independent and the joint PDF assumes non-Boltzmann form.
Despite the fact that for studied anharmonic L\'evy oscillators under nonlinear damping  variances of $x$ and $v$ exist, the  statistical properties of cross-correlation $xv$ cannot be reliably calculated.
Therefore, we have limited ourselves to depicting the sample ratio of probability densities leaving the problem of quantifying the dependence between $x$ and $v$ for further studies.
}


The parabolic addition to the quartic potential destroys the multimodality of overdamped \bdt{steady} states \cite{chechkin2002,chechkin2003,chechkin2004}. 
Therefore, we check the mixture of cubic and linear friction
\begin{equation}
    T(v)=-\gamma ( v^3+a v)\;\;\; (a>0).
    \label{eq:fric4p2}
\end{equation}
Analogously to the overdamped setup \cite{chechkin2002}, increase in $a$ above a critical value $a_c$ ($a_c=0.794$) destroys bimodality in the velocity marginal density $P(v)$.
Moreover, for $a>a_c$, not only $P(v)$ but also the full \bdt{non-equilibrium} stationary density $P(x,v)$ becomes unimodal. 
In contrast, for $0<a<a_c$, the velocity PDF $P(v)$, as well as the full density $P(x,v)$ are multimodal.
At the same time, the marginal position distribution $P(x)$ is unimodal, see Figs.~\ref{fig:X4V4g1} -- \ref{fig:X4V4g6} and \ref{fig:X4V4g4b02} -- \ref{fig:X4V4g6b02}, \bdt{which present results for the Cauchy noise.}

Finally, we have examined  motion of a Langevin particle \bdt{under action of the Cauchy noise} with the velocity dependent friction term given by a polynomial steeper than cubic.
For that purpose we have used the following set of equations
\begin{equation}
\left\{
\begin{array}{lcl}
    \dot{x}(t) & = & v(t) \\
    \dot{v}(t) & = & -\gamma \left[6v^5 - \frac{76}{11}v^3 + 2v\right] - x^3(t) + \zeta(t) 
\end{array}
\right..
\label{eq:nonlinfricV5}
\end{equation}
The friction term was chosen in a such way that, in the absence of $-x^3$ force, the velocity marginal density is trimodal \cite{capala2019multimodal}.
Such a choice of the friction was primarily motivated by possibility of examination of the probability density behavior for the system in which velocity marginal distribution has both zero and non-zero modal values.
Numerical simulations confirm  that, even in the presence of a quartic potential $V(x)$, the velocity marginal distribution remains trimodal. 
Therefore, basing on the marginal velocity distribution, the probability mass or the concentration of particles may be divided into two distinct groups. 
The first group, with velocities corresponding to the outer maxima ($|v|\gg 0$) of the velocity marginal distribution, behaves very similar like particles described by Eq.~(\ref{eq:nonlinfric}).
They produce two modal values corresponding to these velocities. 
At the same time, in the spatial domain, those modes produce a single peak at $x=0$, so that 
the spatial multimodality for $v\neq 0$ is not observed.
The second group of particles includes those ones with the velocity close to zero (represented by the central mode of the velocity marginal distribution).
Due to small velocities, dynamics of particles from the second group can be similar to the overdamped motion. 
For $v \approx 0$, the full probability density $P(x,v)$ has two maxima as 
$P(x,v\approx 0)$ depends nonmonotonously on $x$.
At the same time, the position marginal distribution $P(x)$ remains unimodal, see Fig.~\ref{fig:X4V6g4}, which shows \bdt{NESS} and marginal densities for Eq.~(\ref{eq:nonlinfricV5}) with $\gamma=4$.
Fig.~\ref{fig:X4V6g4} demonstrates the strong trimodality in the marginal \bdt{steady} density $P(v)$, which was already discussed above and in Ref.~\onlinecite{capala2019multimodal}.
The position marginal distribution $P(x)$ stays unimodal, despite of two modal values of the full probability density with modes at  non-zero positions. 
In total, the model described by Eq.~(\ref{eq:nonlinfricV5}) has four modes --- two at $v \approx 0$ and two at $v\neq0$.

The nonlinear friction used in Eq.~(\ref{eq:nonlinfricV5}), i.e.,
 \begin{equation}
     T(v)=-\gamma \left[6v^5 - \frac{76}{11}v^3 + 2v\right].
     \label{eq:nonmonotonousfriction}
 \end{equation}
is a nonmonotonous function of $v$.
Clearly, such a nonlinear friction only dissipate energy, i.e.,  it does not lead to the active L\'evy motion \cite{romanczuk2011brownian,notel2016diffusion}.
In Ref.~\onlinecite{capala2019multimodal} it was shown that, in the overdamped system, the deterministic force given by Eq.~(\ref{eq:nonmonotonousfriction}) with $v$ replaced by $x$ produces trimodal \bdt{non-equilibrium} stationary state.
In accordance with these findings, for the underdamped motion, the velocity distribution is also trimodal with modes located at the same locations as in the underdamped model.
Now, presence of three modes \bdt{in overdamped systems \cite{capala2019multimodal}} can be reinterpreted. 
The most likely values of velocities are those corresponding to the minimal friction. 
Unfortunately, this simple intuitive explanation is of the qualitative type only, because its prediction on the position of modal values are significantly worse than arguments based on the extremes of the  potential curvature \cite{chechkin2002,chechkin2004,capala2019multimodal}.

\section{Summary and Conclusions\label{sec:summary}}

Here we have analyzed numerically stochastic dynamics of underdamped stochastic oscillators subject to velocity-dependent nonlinear damping and additive L\'evy white noise. 

So far, it is known that \bdt{non-equilibrium stationary states (NESS)} for overdamped anharmonic stochastic oscillators, $V(x)=|x|^n/n$, driven by L\'evy noise exist for $n>2-\alpha$. 
More importantly, at $n=2$, the corresponding \bdt{non-equilibrium} stationary PDFs change from unimodal to bimodal forms. 
Emergence of bimodal \bdt{NESS} for $n>2$ can be intuitively explained in the limit of a vanishing noise. 
In the weak noise limit, for $n>2$, time of deterministic sliding from $|x|>0$ to the origin is infinite. 
The competition between deterministic sliding and escapes induced by noise pulses is responsible for depletion of the probability of finding a particle at $x=0$.
In consequence,  $P(x)$ has a local minimum at $x=0$ and the distribution becomes bimodal.
Putting it differently, for overdamped motion in single-well potentials, difficulty in reaching origin is responsible for emergence of bimodal \bdt{NESS}.
The very same scenario is observed for underdamped dynamics.
\bdt{However, in this situation, due to non-zero velocity, a trajectory can more easily visit the origin.
In consequence, it is harder to observe multimimodal non-equilibrium stationary states in underdamped motions than in the overdamped motions.
Moreover, nonlinear friction additionally hampers emergence of multimodal steady states.}

Stochastic underdamped systems are characterized by a velocity and a position, which are distributed according to some probability density.
If the particle moves in the external potential and this movement is subject to damping, the probability density can asymptotically attain the stationary density.
For L\'evy noise, this takes place under the condition that nonlinear friction is strong enough and the potential grows sufficiently fast.
The problem of multimodality of \bdt{NESS} for the underdamped dynamics is more complex than for the overdamped dynamics, because one can ask about multimodality in the full probability density $P(x,v)$ or in marginal \bdt{non-equilibrium} stationary densities $P(x)$ and $P(v)$.

For underdamped motion in the regime of the nonlinear friction it is easy to record multimodality in the velocity, and consequently in the full density, as this feature is mainly determined by the friction term.
At the same time the spatial multimodality is more difficult to be \bdt{induce}.
Importantly, in the underdamped system the \bdt{non-equilibrium} stationary $P(v)$ densities, \bdt{despite action of the additional deterministic $=V'(x)$ force,} are practically the same as $P(x)$ densities  for analogous overdamped systems, i.e., for overdamped systems with the same deterministic force as friction, i.e., $-V'(x)=T(x)$.
For instance, for the linear friction, $P(v)$ densities are of the $\alpha$-stable type.
Nevertheless, some differences between \bdt{non-equilibrium} stationary $P(x)$ in overdamped systems and $P(v)$ in the underdamped system can be produced by the deterministic force $-V'(x)$, which accompanies action of friction in the equation describing time evolution of the velocity.

As the main type of nonlinear friction we have used  $T(v)=-\gamma \sign(v) |v|^{\kappa-1}$, which for $\kappa>2$ is responsible for observation of  multimodal velocity marginal densities $P(v)$.
Nevertheless, the multimodality in $v$ do not transfer into spatial multimodality of \bdt{non-equilibrium stationary states}. 
The increase in damping coefficient $\gamma$ not only influences the modality of \bdt{NESS} but also affects the energy distribution widths, cf. Fig.~\ref{fig:Energia_X4V4g1}.
The width of energy distribution is the decaying function of $\gamma$.
At the same time, the median of kinetic energy is order of magnitude larger than the median of potential energy, because kinetic energy is quadratic in the velocity while the potential energy is quartic function of the position. 
Moreover tails of velocity distribution are heavier than tails of position distribution.
The addition of a linear component to such a friction form, $T(v)=-\gamma \sign(v) |v|^{\kappa-1}-\gamma a v$, is capable of destroying the velocity multimodality, as can be clearly visible for $\kappa=4$, cf. Fig.~\ref{fig:X4V4g6} vs. Fig.~\ref{fig:X4V4g6b02}.
Eventually, higher order damping, e.g., damping given by Eq.~(\ref{eq:nonmonotonousfriction}), produces trimodal \bdt{NESS} in the velocity with one mode at $v\approx 0$ and two modes at $|v|>0$.
On the one hand, particles with $v\approx0$ are close to be overdamped and consequently, due to the potential cubic force, $-V'(x)=-x^3$, they  follow a bimodal distribution.
On the other hand, particles with $|v|>0$ are distributed according to a unimodal density.
The full density $P(x,v)$ has four modal values because for $v=0$ additional spatial multimodality is produced.

Finally, in the limit of $\gamma\to\infty$, velocity becomes overdamped. Actually, already for finite $\gamma$, the motion becomes practically overdamped. Therefore, in the case of linear friction, the \bdt{non-equilibrium} stationary density $P(x)$ approaches the one characterizing overdamped motion in the very same potential.
For the nonlinear friction the situation is very different.
Due to nonlinearity of the damping, overdamped Langevin equation is not restored in the strong friction limit.
Consequently, \bdt{non-equilibrium stationary state} for nonlinear friction is different that the \bdt{steady} state for the overdamped motion in the very same potential.

\section*{Acknowledgement}

This project was supported by the National Science Center (Poland) grant 2018/31/N/ST2/00598.
This research was supported in part by PLGrid Infrastructure.

\section*{Data availability}
The data that support the findings of this study are available from the corresponding author (KC) upon reasonable request.


\end{document}